\documentclass[english]{article}
\usepackage[T1]{fontenc}
\usepackage[latin9]{inputenc}
\usepackage[letterpaper]{geometry}
\usepackage{amstext}
\usepackage[sort&compress]{natbib}
\usepackage{amssymb,amsfonts,amsmath}
\usepackage{epsfig}
\usepackage{amsthm}
\usepackage{amssymb}
\usepackage{amsmath}
\usepackage{graphics}
\usepackage{graphicx}
\usepackage{dcolumn}
\usepackage{bm}
\usepackage{subfigure}
\usepackage{caption}
\usepackage{xcolor}
\usepackage{color}
\usepackage{babel}
\usepackage{setspace}
\usepackage{float}
\usepackage{listings}
\usepackage{multirow}
\usepackage{slashbox}
\usepackage{enumerate}
\usepackage{enumitem}
\setlistdepth{9}

\setlist[itemize,1]{label=\textbullet}
\setlist[itemize,2]{label=\textbullet}
\setlist[itemize,3]{label=\textbullet}
\setlist[itemize,4]{label=\textbullet}
\setlist[itemize,5]{label=\textbullet}
\setlist[itemize,6]{label=\textbullet}
\setlist[itemize,7]{label=\textbullet}
\setlist[itemize,8]{label=\textbullet}
\setlist[itemize,9]{label=\textbullet}

\renewlist{itemize}{itemize}{9}

\usepackage{algorithm}
\usepackage{algorithmic}

\doublespace

\makeatletter

\begin{document}
\title{Bayesian Inference of Gene Expression Dynamics in Alzheimer Brains}
\date{}

\author{Shanjun Mao, Hunan University \\
Xiaodan Fan, The Chinese University of Hong Kong}
\maketitle

\section{Abstract}

Alzheimer's disease (AD) is a serious neurodegenerative disease consisting of four stages where the illness gets progressively worse. It is of great significance to detect the gene regulatory mechanism as AD progresses and, thus, to help us better understand the causes of AD and find ways to treat or control AD. There are numerous researches to conduct this kind of study. However, the majority of methods are processing region by region of brain, stage by stage of AD, and then compare the results to detect changes. It is unclear how to combine these three dimensions, i.e., gene, region and stage, simultaneously to study gene expression dynamics of AD. This is the motivation of our research. In our study, we propose a statistical model of increments to clarify the relationship between gene expression in adjacent stages, so that we could better estimate the missing data we want and obtain a complete reasonable dynamic regulatory network model. Simulations are conducted to validate the statistical power of our algorithm. Moreover, a real data analysis shows that our method can capture the dynamic gene regulatory relationships among this complex brain data.

\section{Introduction}

Alzheimer's disease (AD) is a slow progressing neurodegenerative disease, affecting a lot of people around the world. AD has been identified for more than a century, and although the progression of its pathologic mechanisms still remains unclear, there is increasing interest in its research \citep{Liang2008Neuronal}. Significant advances have been made in understanding the neuropathological, neuroanatomical, and molecular biological abnormalities of the brains of patients with AD \citep{Hardy2006A}. In addition, the advent of high-throughput gene expression techniques enables us to study the expression levels of multiple genes simultaneously to identify AD-related changes in many molecular biological systems \citep{LockhartExpression}.

Most cases of AD have no known genetic, and the underlying genetic modifications are complex and elusive \citep{Bertram2010The}. While transcriptional studies suggest that dysfunctions of cellular pathway may precede neuropathologic markers of AD. Therefore, it is of great significance to study the changes in genetic transcription and cellular pathway of AD. There are numerous researches to conduct this kind of study. \cite{MillerA,Miyashita2014,Shuai2013A} conduct whole-genome microarray analyses through network approaches to determine transcriptional changes and gene co-expression relationships at one stage of AD or among several stages separately. In addition to the genetic changes in the brain, changes among brain regions are also causing concern. As the disease progresses, more and more brain regions are damaged, but AD does not affect all regions simultaneously or consistently \citep{WangFunctional}. Some brain regions are more susceptible to AD than others \citep{Braak1995Staging}. \cite{haroutunian2009transcriptional} identifies the relation between stages of AD and overall changes in gene expression, as well as the brain regions of most susceptible to transcriptional changes as the disease progressed. \cite{WangIntegrative} provides a comprehensive assessment of the critical molecular pathways associated with the pathology of AD by transcriptomic network analysis of 19 brain regions. However, the majority of methods are processing region by region, stage by stage, and then compare the results to detect changes. It is unclear how to combine these three dimensions, i.e., gene, region and stage, simultaneously to study genome-wide gene expression dynamics of AD, and this is our motivation.

\begin{table}
	\centering
	\begin{tabular}{c c c || c | c | c | c}
		& & & Stage 1 & Stage 2 & Stage 3 & Stage 4 \\
		\hline
		\multirow{4}*{Person 1} & \multirow{2}*{Gene 1} & Region 1 & $\checkmark$ & $\bigcirc$ & $\bigcirc$ & $\bigcirc$ \\
		& & Region 2 & $\checkmark$ & $\bigcirc$ & $\bigcirc$ & $\bigcirc$ \\
		& \multirow{2}*{Gene 2} & Region 1 & $\checkmark$ & $\bigcirc$ & $\bigcirc$ & $\bigcirc$ \\
		& & Region 2 & $\checkmark$ & $\bigcirc$ & $\bigcirc$ & $\bigcirc$ \\
		\hline
		\multirow{4}*{Person 2} & \multirow{2}*{Gene 1} & Region 1 & $\times$ & $\checkmark$ & $\bigcirc$ & $\bigcirc$ \\
		& & Region 2 & $\times$ & $\checkmark$ & $\bigcirc$ & $\bigcirc$ \\
		& \multirow{2}*{Gene 2} & Region 1 & $\times$ & $\checkmark$ & $\bigcirc$ & $\bigcirc$ \\
		& & Region 2 & $\times$ & $\checkmark$ & $\bigcirc$ & $\bigcirc$ \\
		\hline
		\multirow{4}*{Person 3} & \multirow{2}*{Gene 1} & Region 1 & $\times$ & $\times$ & $\checkmark$ & $\bigcirc$ \\
		& & Region 2 & $\times$ & $\times$ & $\times$ & $\bigcirc$ \\
		& \multirow{2}*{Gene 2} & Region 1 & $\times$ & $\times$ & $\checkmark$ & $\bigcirc$ \\
		& & Region 2 & $\times$ & $\times$ & $\times$ & $\bigcirc$ \\
	\end{tabular}
	\caption[An example to display characteristics of the brain data]{An example to display characteristics of the brain data. There three symbols, $\checkmark$, $\times$ and $\bigcirc$, represent that the corresponding data is observed, not observed, and not produced, respectively.}
	\label{dataexample}
\end{table}

In addition to considering the above three dimensions at the same time, the high missingness of the brain data is also one of the difficulties of the problem. The characteristic of the brain data is displayed in Table~\ref{dataexample}. The data was observed from brain tissue specimens derived from persons who died at specific stages of AD, thus, for a person, like Person 2 in Table~\ref{dataexample}, we can observe the data at the stage of his death, i.e. Stage 2, but we can not observe its corresponding data at Stage 1 because Person 2 was not dead yet and his brain tissue specimens were not available at Stage 1. Its corresponding data at Stage 3 can't be observed either because the person died before Stage 3. On the other hand, not all brain regions for all persons are available on account of some aliquots from some brain regions which did not yield RNA of sufficient quality or quantity for microarray analysis, like the data of Gene 1 in Region 2 for Person 3 at Stage 3 was not observed.

Our aim is to detect gene regulatory networks of AD with different brain regions during each stage transition. In this study, a principled statistical model for brain data is proposed. A Bayesian approach is presented for estimation. A Markov chain Monte Carlo (MCMC) method, more specifically a Metropolis-within-Partially-collapsed-Gibbs algorithm, is used to draw samples from the posterior.

The remaining content of the chapter is organized as follows: the model of brain data are proposed in Section~\ref{sec:model}; the Bayesian framework is presented in Section~\ref{sec:baye}; main results including simulations and real data analysis are demonstrated in Section~\ref{sec:simul} and \ref{sec:real}, respectively; Section~\ref{sec:con} discusses and concludes this chapter.

\section{Model}\label{sec:model}

In this section, a basic model of brain data is proposed.

Let $D_{g_i r_j,t'}^{e,t}$ be the data of gene $g_i$ in the region $r_j$ at stage $t'$ for the $e$-th person who is observed at stage $t$ and $d_{g_i r_j,t'}^{e,t}$ be its corresponding value, where $t' = 1,\cdots,t$, $t = 1,\cdots,T$, $i = 1,\cdots,G$, $j = 1,\cdots,R$, $e=1,\cdots,n_t$, $n_t$ represents the number of persons who are observed at stage $t$ and $T$, $G$, $R$ are the total number of stages, genes and regions, respectively. If $t'=t$, $D_{g_i r_j,t}^{e,t}$ is the observation, If $t' < t$, $D_{g_i r_j,t'}^{e,t}$ is missing. At the beginning stage, i.e., $t=1$, the expression of all genes in all regions are assumed to follow Gaussian distributions with different expectations:
\[ D_{g_i r_j, 1}^{e,t} \sim N(\mu_{g_i r_j},\sigma_1^2), ~\text{for}~ \left\{ \begin{array}{lr} e=1,\cdots,n_t\\ t=1,\cdots,T \end{array} \right. \]
For $\sigma_1^2$, at the beginning stage of AD, the expression of most genes approach the normal situation, and with the fear of inconsistency, our model assumes that the gene expressions at the beginning stage share the same variance with different expectations. During the transition from stage $t-1$ to stage $t$, denoted as $(t-1,t)$, $t = 2, \cdots, T$, there are three kinds of indicators for each $(g_i, r_j)$: $\gamma_{g_i r_j, t-1, t} = 1$ or $0$ represents $(g_i, r_j)$ is regulated or not, $\beta_{g_i r_j,t-1,t}^{g_m} = 1$ or $0$ represents $(g_i, r_j)$ is regulated by gene $g_m$ or not, $\beta_{g_i r_j,t-1,t}^{g_m r_n} = 1$ or $0$ represents $(g_i, r_j)$ is regulated by gene $g_m$ in region $r_n$, i.e., $(g_m, r_n)$, or not and there are several constraints for them: $\gamma_{g_i r_j, t-1, t} + \sum_{m} \beta_{g_i r_j,t-1,t}^{g_m} = 1$ ,$\gamma_{g_i r_j, t-1, t} + \sum_{n} \beta_{g_i r_j,t-1,t}^{g_m,r_n} = 1$ for $m \ne i$ and $\gamma_{g_i r_j, t-1, t} + \sum_{n \ne j} \beta_{g_i r_j,t-1,t}^{g_m,r_n} = 1$ for $m = i$. Specifically speaking, if $\gamma_{g_i r_j, t-1, t} = 1$, which means $\beta_{g_i r_j,t-1,t}^{g_m} = 0$ and $\beta_{g_i r_j,t-1,t}^{g_m r_n} = 0$ for all other $(g_m, r_n)$, representing gene $g_i$ in the region $r_j$ is not regulated, then, the increment of $(g_i, r_j)$ from stage $t-1$ to stage $t$ follows the Gaussian distribution:
\[ D_{g_i r_j,t}^{e,t'} - D_{g_i r_j,t-1}^{e,t'} \sim N(\mu_2, \sigma_2^2) ~\text{for}~ \left\{ \begin{array}{lr} e=1,\cdots,n_{t'}\\ t=2,\cdots,t'\\ t'=2,\cdots,T \end{array} \right. \]
If there is a unique $(g_m, r_n)$ to make $\beta_{g_i r_j,t-1,t}^{g_m} = 1$ and $\beta_{g_i r_j,t-1,t}^{g_m r_n} = 1$, representing gene $g_i$ in the region $r_j$ is regulated by gene $g_m$ in the region $r_n$, then,
\[ D_{g_i r_j,t}^{e,t'} - D_{g_i r_j,t-1}^{e,t'} | D_{g_m r_n,t-1}^{e,t'} = d_{g_m r_n,t-1}^{e,t'} \sim N(a_{g_i r_j,t-1,t}^{g_m r_n}+b_{g_i r_j,t-1,t}^{g_m r_n} \cdot d_{g_m r_n,t-1}^{e,t'}, \sigma_2^2) ~\text{for}~ \left\{ \begin{array}{lr} e=1,\cdots,n_{t'}\\ t=2,\cdots,t'\\ t'=2,\cdots,T \end{array} \right. \]
The relationship between gene expression changes in adjacent stages is mainly used to find out the existence of regulation. Thus, for simplicity and consistency, our model assumes all increments between adjacent stages follow the same variance, $\sigma_2^2$.

\section{Bayesian Inference}\label{sec:baye}

In this section, a Bayesian framework is proposed to estimate the parameters in the proposed model.

The joint likelihood of the complete data can be expressed as:
\begin{equation*}
	L(\boldsymbol{D} | \boldsymbol{\beta},\boldsymbol{\gamma},\boldsymbol{\mu},\boldsymbol{\sigma},\boldsymbol{a},\boldsymbol{b}) = \prod_{t=1}^{T} \prod_{i=1}^{G} \prod_{j}^{R} \prod_{e=1}^{n_t} p(D_{g_i r_j,1}^{e,t}) \times \prod_{t=2}^{T} \prod_{t'=2}^{t} \prod_{i=1}^{G} \prod_{j}^{R} \prod_{e=1}^{n_{t'}} p(D_{g_i r_j,t'}^{e,t} - D_{g_i r_j,t'-1}^{e,t})
\end{equation*}

\subsection{Prior}

The prior distributions are as follows:
\begin{gather*}
	\mu_{g_i r_j} \sim N(c_{g_i r_j},d_{g_i r_j})\\
	\mu_2 \sim N(c_2,d_2)\\
	\sigma^2_i \sim IG(p_i,q_i), i = 1,2\\
	\gamma_{g_i r_j, t-1, t},\beta_{g_i r_j,t-1,t}^{g_1}, \cdots, \beta_{g_i r_j,t-1,t}^{g_G} \sim multinomial ( \gamma_{g_i r_j, t-1, t} + \sum_{m=1}^R \beta_{g_i r_j,t-1,t}^{g_m} = 1;\boldsymbol{p}=\frac{\boldsymbol{1}_G}{G} )\\
	\left\{ \begin{array}{lr}
		if \ \beta_{g_i r_j,t-1,t}^{g_i} = 1 : \beta_{g_i r_j,t-1,t}^{g_i r_1},\cdots,\beta_{g_i r_j,t-1,t}^{g_i r_{j-1}},\beta_{g_i r_j,t-1,t}^{g_i r_{j+1}},\cdots,\beta_{g_i r_j,t-1,t}^{g_i r_R} \\
		\ \ \ \ \ \ \ \ \ \ \ \ \ \ \ \ \ \ \ \ \ \ \ \sim multinomial ( \sum_{n \ne j} \beta_{g_i r_j,t-1,t}^{g_i r_n} = 1;\boldsymbol{p}=\frac{\boldsymbol{1}_{R-1}}{R-1} ) \\
		if \ \beta_{g_i r_j,t-1,t}^{g_m} = 1, g_m \ne g_i : \beta_{g_i r_j,t-1,t}^{g_m r_1},\cdots,\beta_{g_i r_j,t-1,t}^{g_m r_R} \sim multinomial ( \sum_{n=1}^R \beta_{g_i r_j,t-1,t}^{g_m r_n} = 1;\boldsymbol{p}=\frac{\boldsymbol{1}_{R}}{R} )
	\end{array}
	\right. \\
	(a_{g_i r_j,t-1,t}^{g_m r_n}, b_{g_i r_j,t-1,t}^{g_m r_n})^T | \beta_{g_i r_j,t-1,t}^{g_m r_n} = 1 \sim N(\boldsymbol{\alpha}_{g_i r_j,t-1,t}^{g_m r_n},\sigma^2\boldsymbol{V}_{g_i r_j,t-1,t}^{g_m r_n}) \ and \ \sigma^2 \sim IG(v/2,2v\lambda)
\end{gather*}
where $\boldsymbol{\alpha}_{g_i r_j,t-1,t}^{g_m r_n} = (\alpha_{g_i r_j,t-1,t}^{g_m r_n,a},\alpha_{g_i r_j,t-1,t}^{g_m r_n,b})^T, \boldsymbol{V}_{g_i r_j,t-1,t}^{g_m r_n} = \begin{pmatrix} V_{g_i r_j,t-1,t}^{g_m r_n,a} & 0\\ 0 & v^2/V_{g_i r_j,t-1,t}^{g_m r_n,b} \end{pmatrix}$.

\subsection{Posterior Inference}

In this section, the detailed conditional posterior of all parameters are obtained.

\subsubsection{Missing Data: $D_{g_i r_j,t'}^{e,t}$}

For each observation $D_{g_i r_j,t}^{e,t}$, $t \ge 2$, we estimate its corresponding missing data at all previous stages, $D_{g_i r_j,t'}^{e,t}$, $t' = 1, \cdots, t-1$, and its conditional posterior is as follows.

If $t' = 1$,
\begin{equation*}
	\begin{split}
		p&(D_{g_i r_j,1}^{e,t} | D_{g_i r_j,2}^{e,t}, \boldsymbol{D}_{(-g_i r_j),1}^{e,t}, \boldsymbol{D}_{(-g_i r_j),2}^{e,t},\boldsymbol{\beta},\boldsymbol{\gamma},\boldsymbol{a},\boldsymbol{b},\boldsymbol{\mu},\boldsymbol{\sigma}) \\
		\propto& p(D_{g_i r_j,2}^{e,t} | D_{g_i r_j,1}^{e,t}) p(\boldsymbol{D}_{(-g_i r_j),2}^{e,t} | \boldsymbol{D}_{(-g_i r_j),1}^{e,t}, D_{g_i r_j,1}^{e,t}) p(D_{g_i r_j,1}^{e,t})\\
		=& p(D_{g_i r_j,2}^{e,t} - D_{g_i r_j,1}^{e,t}) p(\boldsymbol{D}_{(-g_i r_j),2}^{e,t} - \boldsymbol{D}_{(-g_i r_j),1}^{e,t} | D_{g_i r_j,1}^{e,t}) p(D_{g_i r_j,1}^{e,t})\\
		\Rightarrow& D_{g_i r_j,1}^{e,t} | D_{g_i r_j,2}^{e,t}, \boldsymbol{D}_{(-g_i r_j),1}^{e,t}, \boldsymbol{D}_{(-g_i r_j),2}^{e,t},\boldsymbol{\beta},\boldsymbol{\gamma},\boldsymbol{a},\boldsymbol{b},\boldsymbol{\mu},\boldsymbol{\sigma} \sim N(\frac{\mu_{miss}}{\tau^2_{miss}},\frac{1}{\tau^2_{miss}})
	\end{split}
\end{equation*}
where $\boldsymbol{D}_{(-g_i r_j),t'}^{e,t}$ represents the complete data at stage $t'$ of all genes in all regions of the $e$-th person observed at stage $t$, except $D_{g_i r_j,t'}^{e,t}$ and\\ $\mu_{miss} = \frac{E(D_{g_i r_j,1}^{e,t})}{\sigma_1^2} + \frac{D_{g_i r_j,2}^{e,t} - E(D_{g_i r_j,2}^{e,t} - D_{g_i r_j,1}^{e,t}) + \sum_{(g_m,r_n) \ne (g_i, r_j)} \beta_{g_m r_n,1,2}^{g_i r_j} \cdot b_{g_m r_n,1,2}^{g_i r_j} (D_{g_m r_n,2}^{e,t} - D_{g_m r_n,1}^{e,t} - a_{g_m r_n,1,2}^{g_i r_j}) }{\sigma_2^2}$, $\tau^2_{miss} = \frac{1}{\sigma_1^2}+\frac{1 + \sum_{(g_m,r_n) \ne (g_i, r_j)} \beta_{g_m r_n,1,2}^{g_i r_j} \cdot (b_{g_m r_n,1,2}^{g_i r_j})^2}{\sigma_2^2}, E(D_{g_i r_j,1}^{e,t}) = \mu_{g_i r_j}$. For $E(D_{g_i r_j,2}^{e,t} - D_{g_i r_j,1}^{e,t})$, during the transition $(t=1,t=2)$, if $(g_i,r_j)$ is not regulated, $E(D_{g_i r_j,2}^{e,t} - D_{g_i r_j,1}^{e,t}) = \mu_2$, if it is regulated by $(g_m, r_n)$, $E(D_{g_i r_j,2}^{e,t} - D_{g_i r_j,1}^{e,t}) = a_{g_i r_j,1,2}^{g_m r_n} + b_{g_i r_j,1,2}^{g_m r_n} \cdot D_{g_m r_n,1}^{e,t}$.

If $t' \ge 2$,
\begin{equation*}
	\begin{split}
		p&(D_{g_i r_j,t'}^{e,t} | D_{g_i r_j,t'+1}^{e,t}, D_{g_i r_j,t'-1}^{e,t}, \boldsymbol{D}_{(-g_i r_j),t'}^{e,t}, \boldsymbol{D}_{(-g_i r_j),t'+1}^{e,t},\boldsymbol{\beta},\boldsymbol{\gamma},\boldsymbol{a},\boldsymbol{b},\boldsymbol{\mu},\boldsymbol{\sigma})\\
		\propto& p(D_{g_i r_j,t'}^{e,t} | D_{g_i r_j,t'-1}^{e,t}) p(D_{g_i r_j,t'+1}^{e,t} | D_{g_i r_j,t'}^{e,t}) p(\boldsymbol{D}_{(-g_i r_j),t'+1}^{e,t} | \boldsymbol{D}_{(-g_i r_j),t'}^{e,t}, D_{g_i r_j,t'}^{e,t})\\
		=& p(D_{g_i r_j,t'}^{e,t} - D_{g_i r_j,t'-1}^{e,t}) p(D_{g_i r_j,t'+1}^{e,t} - D_{g_i r_j,t'}^{e,t}) p(\boldsymbol{D}_{(-g_i r_j),t'+1}^{e,t} - \boldsymbol{D}_{(-g_i r_j),t'}^{e,t} | D_{g_i r_j,t'}^{e,t})\\
		\Rightarrow& D_{g_i r_j,t'}^{e,t} | D_{g_i r_j,t'+1}^{e,t}, D_{g_i r_j,t'-1}^{e,t}, \boldsymbol{D}_{(-g_i r_j),t'}^{e,t}, \boldsymbol{D}_{(-g_i r_j),t'+1}^{e,t},\boldsymbol{\beta},\boldsymbol{\gamma},\boldsymbol{a},\boldsymbol{b},\boldsymbol{\mu},\boldsymbol{\sigma} \sim N(\frac{\mu_{miss}}{\eta_{miss}},\frac{\sigma_2^2}{\eta_{miss}})
	\end{split}
\end{equation*}
where $\eta_{miss} = 2+\sum_{(g_m,r_n) \ne (g_i, r_j)} \beta_{g_m r_n,t',t'+1}^{g_i r_j} \cdot (b_{g_m r_n,t',t'+1}^{g_i r_j})^2$, $\mu_{miss} = (D_{g_i r_j,t'-1}^{e,t} + E(D_{g_i r_j,t'}^{e,t} - D_{g_i r_j,t'-1}^{e,t}))+(D_{g_i r_j,t'+1}^{e,t} - E(D_{g_i r_j,t'+1}^{e,t} - D_{g_i r_j,t'}^{e,t})) + \sum_{(g_m,r_n) \ne (g_i, r_j)} \beta_{g_m r_n,t',t'+1}^{g_i r_j} \cdot b_{g_m r_n,t',t'+1}^{g_i r_j} (D_{g_m r_n,t'+1}^{e,t} - D_{g_m r_n,t'}^{e,t} - a_{g_m r_n,t',t'+1}^{g_i r_j})$. For $E(D_{g_i r_j,t''}^{e,t} - D_{g_i r_j,t''-1}^{e,t}), t'' = t', t'+1$, during the transition $(t''-1,t'')$, if $(g_i,r_j)$ is not regulated, $E(D_{g_i r_j,t''}^{e,t} - D_{g_i r_j,t''-1}^{e,t}) = \mu_2$, if it is regulated by $(g_m, r_n)$, $E(D_{g_i r_j,t''}^{e,t} - D_{g_i r_j,t''-1}^{e,t}) = a_{g_i r_j,t''-1,t''}^{g_m r_n} + b_{g_i r_j,t''-1,t''}^{g_m r_n} \cdot D_{g_m r_n,t''-1}^{e,t}$.

\subsubsection{Parameters: $\mu_{g_i r_j}, \sigma_1^2, \mu_2, \sigma_2^2$}

For the expectations and variations, $\mu_{g_i r_j}, \sigma_1^2, \mu_2, \sigma_2^2$, we have the following conditional posteriors:
\begin{equation*}
	\begin{split}
		p&(\mu_{g_i r_j} | \boldsymbol{D}_{g_i r_j,1}^t, t=1,\cdots, T,\sigma_1^2) \propto \prod_{t=1}^{T} p(\boldsymbol{D}_{g_i r_j,1}^t | \mu_{g_i r_j},\sigma_1^2) p(\mu_{g_i r_j}) \\
		\Rightarrow& \mu_{g_i r_j} | \boldsymbol{D}_{g_i r_j,1}^t, t=1,\cdots, T,\sigma_1^2 \sim N(\frac{\mu_{\mu_{g_i r_j}}}{\tau^2_{\mu_{g_i r_j}}},\frac{1}{\tau^2_{\mu_{g_i r_j}}})
	\end{split}
\end{equation*}
where $\mu_{\mu_{g_i r_j}} = \frac{\sum_{t=1}^{T} \sum_{e=1}^{n_t} D_{g_i r_j,1}^{e,t} }{\sigma_1^2} + \frac{c_{g_i r_j}}{d_{g_i r_j}}$, $\tau^2_{\mu_{g_i r_j}} = \frac{\sum_{t=1}^{T}\sum_{e=1}^{n_t} 1}{\sigma_1^2} + \frac{1}{d_{g_i r_j}}$ and $\boldsymbol{D}_{g_i r_j,t'}^t$ represents the data of gene $i$ in region $j$ at stage $t'$ for all persons who are observed at stage $t$, i.e., $\boldsymbol{D}_{g_i r_j,t'}^t = \{ D_{g_i r_j,t'}^{e,t}; e = 1, \cdots, n_t \}$.

\begin{equation*}
	\begin{split}
		p&(\sigma_1^2 | \boldsymbol{D}_{1}^{t}, t=2,\cdots,T,\boldsymbol{\mu}) \propto \prod_{t=1}^{T} p(\boldsymbol{D}_{1}^{t} | \boldsymbol{\mu},\sigma_1^2) p(\sigma_1^2)\\
		\Rightarrow& \sigma_1^2 | \boldsymbol{D}_{1}^{t}, t=1,\cdots,T,\boldsymbol{\mu} \sim IG(\alpha_{\sigma_1^2},\beta_{\sigma_1^2})
	\end{split}
\end{equation*}
where $\alpha_{\sigma_1^2} = \frac{\sum_{t=1}^{T} \sum_{i=1}^{G} \sum_{j=1}^{R} \sum_{e=1}^{n_t} 1}{2} + p_1$, $\beta_{\sigma_1^2} = \frac{\sum_{t=1}^{T} \sum_{i=1}^{G} \sum_{j=1}^{R} \sum_{e=1}^{n_t} (D_{g_i r_j,1}^{e,t}- \mu_{g_i r_j})^2}{2} + q_1$ and $\boldsymbol{D}_{t'}^{t}$ represents the data of all gene in all region at stage $t'$ for all persons who are observed at stage $t$, i.e., $\boldsymbol{D}_{t'}^{t} = \{ \boldsymbol{D}_{g_i r_j,t'}^t; i = 1, \cdots, G, j = 1, \cdots, R \}$.

\begin{equation*}
	\begin{split}
		p&(\mu_2 | \boldsymbol{D},\boldsymbol{\gamma},\sigma_2^2) \propto p(\boldsymbol{D} | \boldsymbol{\gamma},\mu_2,\sigma_2^2) p(\mu_2) \\
		\Rightarrow& \mu_2 | \boldsymbol{D},\boldsymbol{\gamma},\sigma_2^2 \sim N(\frac{\mu_{\mu_2}}{\tau^2_{\mu_2}},\frac{1}{\tau^2_{\mu_2}})
	\end{split}
\end{equation*}
where $\mu_{\mu_2} = \frac{\sum_{t=2}^{T} \sum_{t'=2}^{t} \sum_{i}^{G} \sum_{j}^{R} \sum_{e}^{n_t} (D_{g_i r_j,t'}^{e,t}-D_{g_i r_j,t'-1}^{e,t}) \cdot \gamma_{g_i r_j,t-1,t}}{\sigma_2^2} + \frac{c_2}{d_2}$, $\tau^2_{\mu_2} = \frac{\sum_{t=2}^{T} \sum_{t'=2}^{t} \sum_{i}^{G} \sum_{j}^{R} \sum_{e}^{n_t} \gamma_{g_i r_j,t-1,t}^2}{\sigma_2^2} + \frac{1}{d_2}$ and $\boldsymbol{D}$ represents the complete data, i.e., $\boldsymbol{D} = \{ \boldsymbol{D}_{t'}^{t}; t' = 1, \cdots, t, t = 1,\cdots,T \}$.

\begin{equation*}
	\begin{split}
		p&(\sigma_2^2 | \boldsymbol{D},\boldsymbol{\beta},\boldsymbol{\gamma},\boldsymbol{a},\boldsymbol{b},\mu_2) \propto p(\boldsymbol{D} | \boldsymbol{\beta},\boldsymbol{\gamma},\boldsymbol{a},\boldsymbol{b},\mu_2,\sigma_2^2) p(\sigma_2^2) \\
		\Rightarrow& \sigma_2^2 | \boldsymbol{D},\boldsymbol{\beta},\boldsymbol{\gamma},\boldsymbol{a},\boldsymbol{b},\mu_2 \sim IG(\alpha_{\sigma_2^2},\beta_{\sigma_2^2})
	\end{split}
\end{equation*}
where $\beta_{\sigma_1^2} = \frac{\sum_{t=2}^{T} \sum_{t'=2}^{t} \sum_{i}^{G} \sum_{j}^{R} \sum_{e}^{n_t} (D_{g_i r_j,t'}^{e,t}-D_{g_i r_j,t'-1}^{e,t}-E(D_{g_i r_j,t'}^{e,t}-D_{g_i r_j,t'-1}^{e,t}))^2}{2} + q_2$, $\alpha_{\sigma_2^2} = \frac{\sum_{t=2}^{T} \sum_{t'=2}^{t} \sum_{i}^{G} \sum_{j}^{R} \sum_{e}^{n_t} 1}{2} + p_2$. For $E(D_{g_i r_j,t'}^{e,t}-D_{g_i r_j,t'-1}^{e,t})$, if $\gamma_{g_i r_j,t'-1,t'} = 1$, $E(D_{g_i r_j,t'}^{e,t}-D_{g_i r_j,t'-1}^{e,t}) = \mu_2$, if $\gamma_{g_i r_j,t'-1,t'} = 0$ and $\beta_{g_i r_j,t'-1,t'}^{g_m r_n} = 1$, $E(D_{g_i r_j,t'}^{e,t}-D_{g_i r_j,t'-1}^{e,t}) = a_{g_i r_j, t'-1, t'}^{g_m r_n}+b_{g_i r_j, t'-1, t'}^{g_m r_n} \cdot D_{g_m r_n,t'-1}^{e,t}$.

\subsubsection{Coefficients: $a_{g_i,r_j,t-1,t}^{g_m,r_n},b_{g_i,r_j,t-1,t}^{g_m,r_n}$}

Given each regulatory relationship, for its corresponding coefficients, we derive its conditional posterior and adopt MH algorithm to get its samples.

The joint prior of all coefficients is:
\begin{equation*}
	\scalebox{0.86}{$
		\begin{split}
			p&(a_{g_i,r_j,t-1,t}^{g_m,r_n},b_{g_i,r_j,t-1,t}^{g_m,r_n}; \text{for all } t, g_i, r_j, g_m, r_n | \boldsymbol{\beta}) \\
			=& \int p(a_{g_i,r_j,t-1,t}^{g_m,r_n},b_{g_i,r_j,t-1,t}^{g_m,r_n}; \text{for all } t, g_i, r_j, g_m, r_n | \boldsymbol{\beta}, \sigma^2) p(\sigma^2) d\sigma^2\\
			=& \int \prod_{t,g_i,r_j,g_m,r_n} p\left( \begin{pmatrix} a_{g_i r_j,t-1,t}^{g_m r_n} \\ b_{g_i r_j,t-1,t}^{g_m r_n} \end{pmatrix} \sim N \begin{pmatrix} \begin{pmatrix} \alpha_{g_i r_j,t-1,t}^{g_m r_n,a} \\ \alpha_{g_i r_j,t-1,t}^{g_m r_n,b} \end{pmatrix}, \sigma^2 \begin{pmatrix} V_{g_i r_j,t-1,t}^{g_m r_n,a} & 0 \\ 0 & v^2/V_{g_i r_j,t-1,t}^{g_m r_n,b} \end{pmatrix} \end{pmatrix} \right)^{\beta_{g_i r_j,t-1,t}^{g_m r_n}} p(\sigma^2 \sim IG(v/2,2v\lambda)) d\sigma^2\\
			\propto& \int (\sigma^2)^{-\sum_{t,g_i,r_j,g_m,r_n} \beta_{g_i r_j,t-1,t}^{g_m r_n}}\\
			\times& \exp \left\{ -\frac{1}{2\sigma^2} \sum_{t,g_i,r_j,g_m,r_n} \beta_{g_i r_j,t-1,t}^{g_m r_n} \left[ \frac{(a_{g_i,r_j,t-1,t}^{g_m,r_n}-\alpha_{g_i r_j,t-1,t}^{g_m r_n,a})^2}{V_{g_i r_j,t-1,t}^{g_m r_n,a}} + \frac{(b_{g_i r_j,t-1,t}^{g_m r_n}-\alpha_{g_i r_j,t-1,t}^{g_m r_n,b})^2}{v^2/V_{g_i r_j,t-1,t}^{g_m r_n,b}} \right] \right\} (\sigma^2)^{-(\frac{v}{2}+1)} e^{-\frac{2v\lambda}{\sigma^2}} d\sigma^2\\
			\propto& \left\{ \sum_{t,g_i,r_j,g_m,r_n} \beta_{g_i r_j,t-1,t}^{g_m r_n} \left[ \frac{(a_{g_i,r_j,t-1,t}^{g_m,r_n}-\alpha_{g_i r_j,t-1,t}^{g_m r_n,a})^2}{V_{g_i r_j,t-1,t}^{g_m r_n,a}} + \frac{(b_{g_i r_j,t-1,t}^{g_m r_n}-\alpha_{g_i r_j,t-1,t}^{g_m r_n,b})^2}{v^2/V_{g_i r_j,t-1,t}^{g_m r_n,b}} \right] + 4v\lambda \right\}^{-(\frac{v}{2}+\sum_{t,g_i,r_j,g_m,r_n} \beta_{g_i r_j,t-1,t}^{g_m r_n})}
		\end{split}
		$}
\end{equation*}

If $\beta_{g_i r_j,t-1,t}^{g_m r_n}=1$, then, we have the following conditional priors of $a_{g_i,r_j,t-1,t}^{g_m,r_n}$ and $b_{g_i,r_j,t-1,t}^{g_m,r_n}$:
\begin{equation*}
	\begin{split}
		\sqrt{\frac{v+2N_{\beta}-1}{A_{g_i r_j,t-1,t}^{g_m r_n} V_{g_i r_j,t-1,t}^{g_m r_n,a}}} \left( a_{g_i r_j,t-1,t}^{g_m r_n} - \alpha_{g_i r_j,t-1,t}^{g_m r_n,a} \right) | \boldsymbol{a}_{-(g_i r_j,t-1,t)}, \boldsymbol{b}, \boldsymbol{\beta} \sim t_{v+2N_{\beta}-1}\\
		\sqrt{\frac{v+2N_{\beta}-1}{B_{g_i r_j,t-1,t}^{g_m r_n} v^2/V_{g_i r_j,t-1,t}^{g_m r_n,b}}} \left( b_{g_i r_j,t-1,t}^{g_m r_n} - \alpha_{g_i r_j,t-1,t}^{g_m r_n,b} \right) | \boldsymbol{b}_{-(g_i r_j,t-1,t)}, \boldsymbol{a}, \boldsymbol{\beta} \sim t_{v+2N_{\beta}-1}
	\end{split}
\end{equation*}
where $\boldsymbol{a}_{-(g_i r_j,t-1,t)}, \boldsymbol{b}_{-(g_i r_j,t-1,t)}$ represents the intercept and scale coefficients of all regulations except for the corresponding one of $(g_i, r_j)$ from $t-1$ to $t$, respectively, $N_{\beta} = \sum_{t,g_i,r_j,g_m,r_n} \beta_{g_i r_j,t-1,t}^{g_m r_n}$, $A_{g_i r_j,t-1,t}^{g_m r_n} = \frac{(b_{g_i r_j,t-1,t}^{g_m r_n}-\alpha_{g_i r_j,t-1,t}^{g_m r_n,b})^2}{v^2/V_{g_i r_j,t-1,t}^{g_m r_n,b}} + 4v\lambda + C_{g_i r_j,t-1,t}^{g_m r_n}$, $B_{g_i r_j,t-1,t}^{g_m r_n} = \frac{(a_{g_i,r_j,t-1,t}^{g_m,r_n}-\alpha_{g_i r_j,t-1,t}^{g_m r_n,a})^2}{V_{g_i r_j,t-1,t}^{g_m r_n,a}} + 4v\lambda + C_{g_i r_j,t-1,t}^{g_m r_n}$, and
\begin{equation*}
	\begin{split}
		C_{g_i r_j,t-1,t}^{g_m r_n} = \sum_{t,(g_{i'},r_{j'})\ne(g_i,r_j),g_{m'},r_{n'}} \beta_{g_{i'},r_{j'},t-1,t}^{g_{m'},r_{n'}} \left[ \frac{(a_{g_{i'},r_{j'},t-1,t}^{g_{m'},r_{n'}}-\alpha_{g_{i'} r_{j'},t-1,t}^{g_{m'} r_{n'},a})^2}{V_{g_{i'} r_{j'},t-1,t}^{g_{m'} r_{n'},a}} + \frac{(b_{g_{i'} r_{j'},t-1,t}^{g_{m'} r_{n'}}-\alpha_{g_{i'} r_{j'},t-1,t}^{g_{m'} r_{n'},b})^2}{v^2/V_{g_{i'} r_{j'},t-1,t}^{g_{m'} r_{n'},b}} \right]
	\end{split}
\end{equation*}

Then, we can get the conditional posteriors of coefficients $a_{g_i,r_j,t-1,t}^{g_m,r_n},b_{g_i,r_j,t-1,t}^{g_m,r_n}$:
\begin{equation*}
	\begin{split}
		& p(a_{g_i r_j, t-1, t}^{g_m r_n} | \boldsymbol{D}_{g_i r_j,t}^{t'}, \boldsymbol{D}_{g_i r_j,t-1}^{t'}, \boldsymbol{D}_{g_m r_n,t-1}^{t'}, t'=t,\cdots,T, b_{g_i r_j, t-1, t}^{g_m r_n}, \sigma_2^2)\\
		& \propto p\left\{ \sqrt{\frac{v+2N_{\beta}-1}{A_{g_i r_j,t-1,t}^{g_m r_n} V_{g_i r_j,t-1,t}^{g_m r_n,a}}} \left( a_{g_i r_j,t-1,t}^{g_m r_n} - \alpha_{g_i r_j,t-1,t}^{g_m r_n,a} \right) \sim t_{v+2N_{\beta}-1} \right\}\\
		& \times p\left\{ a_{g_i r_j, t-1, t}^{g_m r_n} \sim N(\frac{\sum_{t'=t}^{T}\sum_{e=1}^{n_{t'}} (D_{g_i r_j,t}^{e,t'}-D_{g_i r_j,t-1}^{e,t'} - b_{g_i r_j, t-1, t}^{g_m r_n} \cdot D_{g_m r_n,t-1}^{e,t'})}{\sum_{t'=t}^{T}\sum_{e=1}^{n_{t'}} 1},\frac{\sigma_2^2}{\sum_{t'=t}^{T}\sum_{e=1}^{n_{t'}} 1}) \right\} \\
		& p(b_{g_i r_j, t-1, t}^{g_m r_n} | \boldsymbol{D}_{g_i r_j,t}^{t'}, \boldsymbol{D}_{g_i r_j,t-1}^{t'}, \boldsymbol{D}_{g_m r_n,t-1}^{t'}, t'=t,\cdots,T, a_{g_i r_j, t-1, t}^{g_m r_n}, \sigma_2^2)\\
		& \propto p\left\{ \sqrt{\frac{v+2N_{\beta}-1}{B_{g_i r_j,t-1,t}^{g_m r_n} v^2/V_{g_i r_j,t-1,t}^{g_m r_n,b}}} \left( b_{g_i r_j,t-1,t}^{g_m r_n} - \alpha_{g_i r_j,t-1,t}^{g_m r_n,b} \right) \sim t_{v+2N_{\beta}-1} \right\}\\
		& \times p\left\{ b_{g_i r_j, t-1, t}^{g_m r_n} \sim N(\frac{\sum_{t'=t}^{T}\sum_{e=1}^{n_{t'}} (D_{g_i r_j,t}^{e,t'}-D_{g_i r_j,t-1}^{e,t'} - a_{g_i r_j, t-1, t}^{g_m r_n}) \cdot D_{g_m r_n,t-1}^{e,t'}}{\sum_{t'=t}^{T}\sum_{e=1}^{n_{t'}} (D_{g_m r_n,t-1}^{e,t'})^2},\frac{\sigma_2^2}{\sum_{t'=t}^{T}\sum_{e=1}^{n_{t'}} (D_{g_m r_n,t-1}^{e,t'})^2}) \right\}
	\end{split}
\end{equation*}

\subsubsection{Regulatory Relationship}

For the regulatory relationship of $(g_i, r_j)$ during the transition $(t-1, t)$, denoted as $m^{t-1,t}_{g_i, r_j}$, there are $G \times R$ situations, and only only situation will happen. Thus, this is going to involve the change in the dimension of its corresponding coefficients $a_{g_i r_j, t-1, t}^{g_m r_n}, b_{g_i r_j, t-1, t}^{g_m r_n}$. We adopt Partially Collapsed Gibbs (PCG) sampler, which means that when we sample $m^{t-1,t}_{g_i, r_j}$, its coefficients $a_{g_i r_j, t-1, t}^{g_m r_n}, b_{g_i r_j, t-1, t}^{g_m r_n}$ are collapsed. Here is a simple example with three stages to elaborate PCG sampler:
\begin{itemize}
	\item[A.] Parent Gibbs sampler
	\begin{itemize}
		\item[(1)] p($\boldsymbol{m}^{1,2}~|~\boldsymbol{D}_{t'}^t,t'=1,2, t=2,\cdots,T, \boldsymbol{a}^{1,2},\boldsymbol{b}^{1,2},\boldsymbol{\mu},\boldsymbol{\sigma}^2$)
		\item[(2)] p($\boldsymbol{a}^{1,2},\boldsymbol{b}^{1,2}~|~\boldsymbol{m}^{1,1},\boldsymbol{D}_{t'}^t,t'=1,2, t=2,\cdots,T,\boldsymbol{\mu},\boldsymbol{\sigma}^2$)
		\item[(3)] p($\boldsymbol{D}_1^t,~|~\boldsymbol{m}^{1,2},\boldsymbol{D}_2^t,\boldsymbol{a}^{1,2},\boldsymbol{b}^{1,2},\boldsymbol{\mu},\boldsymbol{\sigma}^2$) for $t=2,\cdots,T$
		\item[(4)] p($\boldsymbol{m}^{2,3}~|~\boldsymbol{D}_{t'}^t,t'=2,3, t=3,\cdots,T, \boldsymbol{a}^{2,3},\boldsymbol{b}^{2,3},\boldsymbol{\mu},\boldsymbol{\sigma}^2$)
		\item[(5)] p($\boldsymbol{a}^{2,3},\boldsymbol{b}^{2,3}~|~\boldsymbol{m}^{2,3},\boldsymbol{D}_{t'}^t,t'=2,3, t=2,\cdots,T,\boldsymbol{\mu},\boldsymbol{\sigma}^2$)
		\item[(6)] p($\boldsymbol{D}_2^{t}~|~\boldsymbol{m}^{1,2},\boldsymbol{m}^{2,3}, \boldsymbol{D}_1^{t},\boldsymbol{D}_3^{t},\boldsymbol{a}^{2,3},\boldsymbol{b}^{2,3},\boldsymbol{\mu},\boldsymbol{\sigma}^2$) for $t=3,\cdots,T$
		\item[(7)] p($\boldsymbol{\mu},\boldsymbol{\sigma}^2~|~\boldsymbol{m},\boldsymbol{ab},\boldsymbol{\mu},\boldsymbol{\sigma}^2,\boldsymbol{D}$)
	\end{itemize}
	\item[B.] Partially Collapsed Gibbs sampler
	\begin{itemize}
		\item[(1)] p($\boldsymbol{m}^{1,2}~|~\boldsymbol{D}_{t'}^t,t'=1,2, t=2,\cdots,T,\boldsymbol{\mu},\boldsymbol{\sigma}^2$)
		\item[(2)] p($\boldsymbol{a}^{1,2},\boldsymbol{b}^{1,2}~|~\boldsymbol{m}^{1,1},\boldsymbol{D}_{t'}^t,t'=1,2, t=2,\cdots,T,\boldsymbol{\mu},\boldsymbol{\sigma}^2$)
		\item[(3)] p($\boldsymbol{D}_1^t,~|~\boldsymbol{m}^{1,2},\boldsymbol{D}_2^t,\boldsymbol{a}^{1,2},\boldsymbol{b}^{1,2},\boldsymbol{\mu},\boldsymbol{\sigma}^2$) for $t=2,\cdots,T$
		\item[(4)] p($\boldsymbol{m}^{2,3}~|~\boldsymbol{D}_{t'}^t,t'=2,3, t=3,\cdots,T,\boldsymbol{\mu},\boldsymbol{\sigma}^2$)
		\item[(5)] p($\boldsymbol{a}^{2,3},\boldsymbol{b}^{2,3}~|~\boldsymbol{m}^{2,3},\boldsymbol{D}_{t'}^t,t'=2,3, t=2,\cdots,T,\boldsymbol{\mu},\boldsymbol{\sigma}^2$)
		\item[(6)] p($\boldsymbol{D}_2^{t}~|~\boldsymbol{m}^{1,2},\boldsymbol{m}^{2,3}, \boldsymbol{D}_1^{t},\boldsymbol{D}_3^{t},\boldsymbol{a}^{2,3},\boldsymbol{b}^{2,3},\boldsymbol{\mu},\boldsymbol{\sigma}^2$) for $t=3,\cdots,T$
		\item[(7)] p($\boldsymbol{\mu},\boldsymbol{\sigma}^2~|~\boldsymbol{m},\boldsymbol{ab},\boldsymbol{\mu},\boldsymbol{\sigma}^2,\boldsymbol{D}$)
	\end{itemize}
\end{itemize}

We can find that the PCG sampler avoids the use of reversible jump MCMC algorithm by marginalizing over the coefficient whose dimension is not fixed. For the inference of indicators $\gamma_{g_i r_j,t-1,t}$ and $\beta_{g_i r_j,t-1,t}^{g_m r_n}$, let's get the conditional posterior of $\sigma^2$ first:
\begin{equation*}
	\begin{split}
		p&(\sigma^2 | \boldsymbol{a}_{-(g_i r_j,t-1,t)}, \boldsymbol{b}_{-(g_i r_j,t-1,t)}, \boldsymbol{\beta}) \propto p(\boldsymbol{a}_{-(g_i r_j,t-1,t)}, \boldsymbol{b}_{-(g_i r_j,t-1,t)} | \boldsymbol{\beta}, \sigma^2) p(\sigma^2)\\
		\propto& (\sigma^2)^{-(\frac{v}{2}+1)} e^{-\frac{2v\lambda}{\sigma^2}} (\sigma^2)^{-\sum_{t,(g_{i'},r_{j'})\ne(g_i,r_j),g_{m'},r_{n'}} \beta_{g_{i'},r_{j'},t-1,t}^{g_{m'},r_{n'}}}\\
		\times& \exp \left\{ -\frac{1}{2\sigma^2} \sum_{t,(g_{i'},r_{j'})\ne(g_i,r_j),g_{m'},r_{n'}} \beta_{g_{i'},r_{j'},t-1,t}^{g_{m'},r_{n'}} \left[ \frac{(a_{g_{i'},r_{j'},t-1,t}^{g_{m'},r_{n'}}-\alpha_{g_{i'} r_{j'},t-1,t}^{g_{m'} r_{n'},a})^2}{V_{g_{i'} r_{j'},t-1,t}^{g_{m'} r_{n'},a}} + \frac{(b_{g_{i'} r_{j'},t-1,t}^{g_{m'} r_{n'}}-\alpha_{g_{i'} r_{j'},t-1,t}^{g_{m'} r_{n'},b})^2}{v^2/V_{g_{i'} r_{j'},t-1,t}^{g_{m'} r_{n'},b}} \right] \right\}\\
		\Rightarrow & \sigma^2 | \boldsymbol{a}_{-(g_i r_j,t-1,t)}, \boldsymbol{b}_{-(g_i r_j,t-1,t)}, \boldsymbol{\beta} \sim IG(\frac{v}{2}+N_{\beta, g_i r_j,t-1,t}^{g_m r_n}, 2v\lambda+\frac{1}{2}C_{g_i r_j,t-1,t}^{g_m r_n})
	\end{split}
\end{equation*}
where $N_{\beta, g_i r_j,t-1,t}^{g_m r_n} = \sum_{t,(g_{i'},r_{j'})\ne(g_i,r_j),g_{m'},r_{n'}} \beta_{g_{i'},r_{j'},t-1,t}^{g_{m'},r_{n'}}$.

Thus, for the data $\boldsymbol{D}_{g_i r_j,t}^{t'}, t' \ge t$, we have:
\begin{equation*}
	\scalebox{0.95}{$
		\begin{split}
			& p(\boldsymbol{D}_{g_i r_j,t}^{t'}, \boldsymbol{D}_{g_i r_j,t-1}^{t'}, t' \ge t | \gamma_{g_i r_j,t-1,t} = 1, \mu_2, \sigma_2^2) \propto \prod_{t'=t}^{T} \prod_{e=1}^{n_{t'}} \frac{1}{\sqrt{2\pi\sigma_2^2}} \exp \left\{ -\frac{ (D_{g_i r_j,t}^{e,t'}-D_{g_i r_j,t-1}^{e,t'} - \mu_2)^2}{2 \sigma_2^2} \right\}\\
			& p(\boldsymbol{D}_{g_i r_j,t}^{t'}, \boldsymbol{D}_{g_i r_j,t-1}^{t'}, t' \ge t | \beta_{g_i r_j,t-1,t}^{g_m r_n} = 1,\sigma_2^2, \boldsymbol{a}_{-(g_i r_j,t-1,t)}, \boldsymbol{b}_{-(g_i r_j,t-1,t)})\\
			& \propto \int \int \int \prod_{t'=t}^{T} p(\boldsymbol{D}_{g_i r_j,t}^{t'} - \boldsymbol{D}_{g_i r_j,t-1}^{t'} | \beta_{g_i r_j,t-1,t}^{g_m r_n} = 1,\sigma_2^2,a_{g_i r_j, t-1, t}^{g_m r_n}, b_{g_i r_j, t-1, t}^{g_m r_n}) p(a_{g_i r_j, t-1, t}^{g_m r_n}, b_{g_i r_j, t-1, t}^{g_m r_n} | \sigma^2)\\
			& p(\sigma^2 | \boldsymbol{a}_{-(g_i r_j,t-1,t)}, \boldsymbol{b}_{-(g_i r_j,t-1,t)})d a_{g_i r_j, t-1, t}^{g_m r_n} d b_{g_i r_j, t-1, t}^{g_m r_n} d \sigma^2\\
			& = \frac{\Gamma ((v+2N_{\beta, g_i r_j,t-1,t}^{g_m r_n}+\sum_{t'=t}^{T} n_t)/2)(4v\lambda+C_{\beta, g_i r_j,t-1,t}^{g_m r_n})^{(v+2N_{\beta, g_i r_j,t-1,t}^{g_m r_n})/2}}{\pi^{\sum_{t'=t}^{T} n_t/2} |\sigma_2^2\boldsymbol{I} + \boldsymbol{X}_{g_m r_n,t-1}^{(t-T)} \boldsymbol{V}_{g_i r_j,t-1,t}^{g_m r_n} (\boldsymbol{X}_{g_m r_n,t-1}^{(t-T)})^T|^{1/2}}\\
			& \times \left[ 4v\lambda+C_{\beta, g_i r_j,t-1,t}^{g_m r_n} + (\boldsymbol{D}_{g_i r_j,t}^{(t-T)} - \boldsymbol{D}_{g_i r_j,t-1}^{(t-T)} - \boldsymbol{X}_{g_m r_n,t-1}^{(t-T)} \boldsymbol{\alpha}_{g_i r_j,t-1,t}^{g_m r_n})^T(\sigma_2^2\boldsymbol{I} + \boldsymbol{X}_{g_m r_n,t-1}^{(t-T)} \boldsymbol{V}_{g_i r_j,t-1,t}^{g_m r_n} (\boldsymbol{X}_{g_m r_n,t-1}^{(t-T)})^T)^{-1} \right. \\
			& \left. (\boldsymbol{D}_{g_i r_j,t}^{(t-T)} - \boldsymbol{D}_{g_i r_j,t-1}^{(t-T)} - \boldsymbol{X}_{g_m r_n,t-1}^{(t-T)} \boldsymbol{\alpha}_{g_i r_j,t-1,t}^{g_m r_n}) \right]^{-(v+2N_{\beta, g_i r_j,t-1,t}^{g_m r_n}+\sum_{t'=t}^{T} n_t)/2}\\
			& \Rightarrow \boldsymbol{D}_{g_i r_j,t}^{(t-T)} - \boldsymbol{D}_{g_i r_j,t-1}^{(t-T)} | \beta_{g_i r_j,t-1,t}^{g_m r_n} = 1, \sigma_2^2, \boldsymbol{a}_{-(g_i r_j,t-1,t)}, \boldsymbol{b}_{-(g_i r_j,t-1,t)} \sim t_{v+2N_{\beta, g_i r_j,t-1,t}^{g_m r_n}}(\boldsymbol{\mu}, \boldsymbol{\Sigma})\\
			& p(\boldsymbol{D}_{g_i r_j,t}^{t'}, \boldsymbol{D}_{g_i r_j,t-1}^{t'}, t' \ge t | \beta_{g_i r_j,t-1,t}^{g_m} = 1, \sigma_2^2, \boldsymbol{a}_{-(g_i r_j,t-1,t)}, \boldsymbol{b}_{-(g_i r_j,t-1,t)})\\
			& = \sum_{r_n} p(\boldsymbol{D}_{g_i r_j,t}^{t'}, \boldsymbol{D}_{g_i r_j,t-1}^{t'}, t' \ge t | \beta_{g_i r_j,t-1,t}^{g_m r_n} = 1,\sigma_2^2, \boldsymbol{a}_{-(g_i r_j,t-1,t)}, \boldsymbol{b}_{-(g_i r_j,t-1,t)}) p(\beta_{g_i r_j,t-1,t}^{g_m r_n} = 1 | \beta_{g_i r_j,t-1,t}^{g_m} = 1)
		\end{split}
		$}
\end{equation*}
where $\boldsymbol{\mu}=\boldsymbol{X}_{g_m r_n,t-1}^{(t-T)} \boldsymbol{\alpha}_{g_i r_j,t-1,t}^{g_m r_n}$,$\boldsymbol{\Sigma}=\frac{4v\lambda+C_{\beta, g_i r_j,t-1,t}^{g_m r_n}}{v+2N_{\beta, g_i r_j,t-1,t}^{g_m r_n}}(\sigma_2^2\boldsymbol{I} + \boldsymbol{X}_{g_m r_n,t-1}^{(t-T)} \boldsymbol{V}_{g_i r_j,t-1,t}^{g_m r_n} (\boldsymbol{X}_{g_m r_n,t-1}^{(t-T)})^T)$, $\boldsymbol{X}_{g_m r_n,t-1}^{(t-T)} = (\boldsymbol{1},\boldsymbol{D}_{g_m r_n,t-1}^{(t-T)})$, and $\boldsymbol{D}_{g_i r_j,t'}^{(t-T)} = \begin{pmatrix} \boldsymbol{D}_{g_i r_j,t'}^{(t)} \\ \boldsymbol{D}_{g_i r_j,t'}^{(t+1)} \\ \cdots \\ \boldsymbol{D}_{g_i r_j,t'}^{(T)} \end{pmatrix}$.

Thus, for $(g_i, r_j)$, its parameters $\gamma_{g_i r_j,t-1,t}, \beta_{g_i r_j,t-1,t}^{g_m}, \beta_{g_i r_j,t-1,t}^{g_m r_n}$ satisfy:
\begin{equation*}
	\scalebox{0.88}{$
		\begin{split}
			& p(\gamma_{g_i r_j,t-1,t} = 1 | \boldsymbol{D}_{g_i r_j,t}^{t'}, \boldsymbol{D}_{g_i r_j,t-1}^{t'}, t' \ge t, \boldsymbol{\mu}, \boldsymbol{\sigma}, \boldsymbol{a}, \boldsymbol{b}) \propto p(\gamma_{g_i r_j,t-1,t} = 1) p(\boldsymbol{D}_{g_i r_j,t}^{t'} - \boldsymbol{D}_{g_i r_j,t-1}^{t'}, t' \ge t | \gamma_{g_i r_j,t-1,t} = 1, \boldsymbol{\mu}, \boldsymbol{\sigma}, \boldsymbol{a}, \boldsymbol{b})\\
			& p(\beta_{g_i r_j,t-1,t}^{g_m} = 1 | \boldsymbol{D}_{g_i r_j,t}^{t'}, \boldsymbol{D}_{g_i r_j,t-1}^{t'}, t' \ge t, \boldsymbol{\mu}, \boldsymbol{\sigma}, \boldsymbol{a}, \boldsymbol{b}) \propto p(\beta_{g_i r_j,t-1,t}^{g_m} = 1)p(\boldsymbol{D}_{g_i r_j,t}^{t'} - \boldsymbol{D}_{g_i r_j,t-1}^{t'}, t' \ge t | \beta_{g_i r_j,t-1,t}^{g_m} = 1, \boldsymbol{\mu}, \boldsymbol{\sigma}, \boldsymbol{a}, \boldsymbol{b})\\
			& p(\beta_{g_i r_j,t-1,t}^{g_m r_n} = 1 | \boldsymbol{D}_{g_i r_j,t}^{t'}, \boldsymbol{D}_{g_i r_j,t-1}^{t'}, t' \ge t, \boldsymbol{\mu}, \boldsymbol{\sigma}, \boldsymbol{a}, \boldsymbol{b})\\
			&\propto p(\beta_{g_i r_j,t-1,t}^{g_m r_n} = 1 | \beta_{g_i r_j,t-1,t}^{g_m} = 1)p(\boldsymbol{D}_{g_i r_j,t}^{t'} - \boldsymbol{D}_{g_i r_j,t-1}^{t'}, t' \ge t | \beta_{g_i r_j,t-1,t}^{g_m r_n} = 1,\boldsymbol{\mu}, \boldsymbol{\sigma}, \boldsymbol{a}, \boldsymbol{b})
		\end{split}
		$}
\end{equation*}

\subsection{Algorithm}
The basic steps of the whole algorithm are shown in Algorithm~\ref{algorithm}.
\begin{algorithm}
	\caption{Algorithm for Brain Data}
	\label{algorithm}
	\begin{algorithmic}[1]
		\REQUIRE ~~\\ 
		The observation $\boldsymbol{D}_{t}^{t}, t=1,\cdots,T$;\\
		The initial regulatory relationships $\boldsymbol{M}^{t-1,t}_0, t=2,\cdots,T$
		\ENSURE ~~\\ 
		The new regulatory relationships during each stage transition $\boldsymbol{M}^{t-1,t}, t=2,\cdots,T$;\\
		The missing data $\boldsymbol{D}_{t'}^{t}, t'=1,\cdots,t, t=2,\cdots,T$;\\
		All other parameters $\boldsymbol{\mu}, \boldsymbol{\sigma}$;
		
		\FOR{$t=2;t\le T; t++$}
		\FOR{$l=1;l\le L; l++$}
		\STATE propose a candidate model $\boldsymbol{M}^{t-1,t}_* \sim g(\boldsymbol{M} | \boldsymbol{M}^{t-1,t}_{l-1})$ according to some proposal distribution g
		\STATE calculate acceptance rate: $\theta = \min (\frac{p(\boldsymbol{M}^{t-1,t}_*|\boldsymbol{D})}{p(\boldsymbol{M}^{t-1,t}_{l-1}|\boldsymbol{D})} \cdot \frac{g(\boldsymbol{M}^{t-1,t}_{l-1}|\boldsymbol{M}^{t-1,t}_*)}{g(\boldsymbol{M}^{t-1,t}_*|\boldsymbol{M}^{t-1,t}_{l-1})},1)$
		\STATE sample $u \sim U(0,1)$
		\IF{$u < \theta$}
		\STATE accept the proposal $\boldsymbol{M}^{t-1,t}_l = \boldsymbol{M}^{t-1,t}_*$
		\ELSE
		\STATE Reject the proposal $\boldsymbol{M}^{t-1,t}_l = \boldsymbol{M}^{t-1,t}_{l-1}$
		\ENDIF
		\STATE update $\boldsymbol{a},\boldsymbol{b}$
		\ENDFOR
		\STATE let $\boldsymbol{M}^{t-1,t} = \boldsymbol{M}^{t-1,t}_L$
		\STATE sample the missing data $\boldsymbol{D}_{t-1}^{t}~|~\boldsymbol{M}^{t'-1,t'},\boldsymbol{D}_{t''}^{t'}, t'' \le t'-2, t' \le T,\boldsymbol{a},\boldsymbol{b}, \boldsymbol{\mu}, \boldsymbol{\sigma^2}$
		\ENDFOR
		\STATE sample $(\boldsymbol{\mu}, \boldsymbol{\sigma^2})~|~\boldsymbol{M},\boldsymbol{D},\boldsymbol{a},\boldsymbol{b}$
	\end{algorithmic}
\end{algorithm}

For the production of the proposal model $\boldsymbol{M}^{t-1,t}_*$ based on a given model $\boldsymbol{M}^{t-1,t}_{l-1}$, there are three alternative operations to conduct which are described in Algorithm~\ref{operation_add},\ref{operation_delete},\ref{operation_swap}. Moreover, to show the connection between regions for the same gene, when we sample a different $(g_m,r_n)$ for $(g_i,r_j)$, we are going to sample $g_m$ first, and then sample $r_n$.
\begin{algorithm}
	\caption{Adding a regulatory relationship (Add)}
	\label{operation_add}
	\begin{algorithmic}[1]
		\REQUIRE ~~\\ 
		A regulatory model $\boldsymbol{M}^{t-1,t}$ with at least one gene in one region which is not regulated
		\ENSURE ~~\\ 
		A new regulatory mode $\boldsymbol{M}^{t-1,t}_*$
		
		\STATE find all genes and regions which are not regulated in $\boldsymbol{M}^{t-1,t}$, denoted as set $\{ (g_i',r_j') \}$ with $\gamma_{g_i' r_j',t-1,t}=1$
		\STATE calculate the posterior probability $p(\gamma_{g_i' r_j',t-1,t}=1 | \boldsymbol{D},\boldsymbol{\mu},\boldsymbol{\sigma},\boldsymbol{b},\boldsymbol{a})$
		\STATE sample a $(g_i*,r_j*)$ in set $\{ (g_i',r_j') \}$ with the probability proportional to $1-p(\gamma_{g_i' r_j',t-1,t}=1 | \boldsymbol{D},\boldsymbol{\mu},\boldsymbol{\sigma},\boldsymbol{b},\boldsymbol{a})$
		\STATE calculate the posterior probability $p(\beta_{g_i* r_j*,t-1,t}^{g_m' r_n'}=1 | \boldsymbol{D},\boldsymbol{\mu},\boldsymbol{\sigma},\boldsymbol{b},\boldsymbol{a})$ for all $(g_m',r_n')\ne (g_i*,r_j*)$
		\STATE sample a $(g_m*,r_n*)$ in set $\{ (g_m',r_n')\ne (g_i*,r_j*) \}$ with the probability proportional to $p(\beta_{g_i* r_j*,t-1,t}^{g_m' r_n'}=1 | \boldsymbol{D},\boldsymbol{\mu},\boldsymbol{\sigma},\boldsymbol{b},\boldsymbol{a})$
		\STATE generate a new model $\boldsymbol{M}^{t-1,t}_*$ by making $\beta_{g_i* r_j*,t-1,t}^{g_m* r_n*}=1$
	\end{algorithmic}
\end{algorithm}

\begin{algorithm}
	\caption{Deleting a regulatory relationship (Delete)}
	\label{operation_delete}
	\begin{algorithmic}[1]
		\REQUIRE ~~\\ 
		A regulatory model $\boldsymbol{M}^{t-1,t}$ with at least one regulatory relationship
		\ENSURE ~~\\ 
		A new regulatory mode $\boldsymbol{M}^{t-1,t}_*$
		
		\STATE find all regulatory relationships in $\boldsymbol{M}^{t-1,t}$ denoted as set $\{ (g_i',r_j',g_m',r_n') \}$ which means $\beta_{g_i' r_j',t-1,t}^{g_m' r_n'}=1$
		\STATE calculate the posterior probability $p(\beta_{g_i' r_j',t-1,t}^{g_m' r_n'}=1 | \boldsymbol{D},\boldsymbol{\mu},\boldsymbol{\sigma},\boldsymbol{b},\boldsymbol{a})$
		\STATE sample a $(g_i*,r_j*,g_m*,r_n*)$ in set $\{ (g_i',r_j',g_m',r_n') \}$ with the probability proportional to $1-p(\beta_{g_i* r_j*,t-1,t}^{g_m* r_n*}=1 | \boldsymbol{D},\boldsymbol{\mu},\boldsymbol{\sigma},\boldsymbol{b},\boldsymbol{a})$
		\STATE generate a new model $\boldsymbol{M}^{t-1,t}_*$ by making $\gamma_{g_i* r_j*,t-1,t} = 1$
	\end{algorithmic}
\end{algorithm}

\begin{algorithm}
	\caption{Swapping a regulatory relationship (Swap)}
	\label{operation_swap}
	\begin{algorithmic}[1]
		\REQUIRE ~~\\ 
		A regulatory model $\boldsymbol{M}^{t-1,t}$
		\ENSURE ~~\\ 
		A new regulatory mode $\boldsymbol{M}^{t-1,t}_*$
		
		\STATE find all regulatory relationships in $\boldsymbol{M}^{t-1,t}$ denoted as set $\{ (g_i',r_j',g_m',r_n') \}$ which means $\beta_{g_i' r_j',t-1,t}^{g_m' r_n'}=1$
		\STATE calculate the posterior probability $p(\beta_{g_i' r_j',t-1,t}^{g_m' r_n'}=1 | \boldsymbol{D},\boldsymbol{\mu},\boldsymbol{\sigma},\boldsymbol{b},\boldsymbol{a})$
		\STATE sample a $(g_i*,r_j*,g_m*,r_n*)$ in set $\{ (g_i',r_j',g_m',r_n') \}$ with the probability proportional to $1-p(\beta_{g_i* r_j*,t-1,t}^{g_m* r_n*}=1 | \boldsymbol{D},\boldsymbol{\mu},\boldsymbol{\sigma},\boldsymbol{b},\boldsymbol{a})$
		\STATE calculate the posterior probability $p(\beta_{g_i* r_j*,t-1,t}^{g_m'' r_n''}=1 | \boldsymbol{D},\boldsymbol{\mu},\boldsymbol{\sigma},\boldsymbol{b},\boldsymbol{a})$ for all $(g_m'',r_n'')\ne (g_i*,r_j*)$
		\STATE sample a $(g_m**,r_n**)$ in set $\{ (g_m'',r_n'')\ne (g_i*,r_j*) \text{ and } (g_m*,r_j*) \}$ with the probability proportional to $p(\beta_{g_i* r_j*,t-1,t}^{g_m'' r_n''}=1 | \boldsymbol{D},\boldsymbol{\mu},\boldsymbol{\sigma},\boldsymbol{b},\boldsymbol{a})$
		\STATE generate a new model $\boldsymbol{M}^{t-1,t}_*$ by making $\beta_{g_i* r_j*,t-1,t}^{g_m* r_n*} = 0$ and $\beta_{g_i* r_j*,t-1,t}^{g_m** r_n**} = 1$
	\end{algorithmic}
\end{algorithm}

\section{Simulation Results}\label{sec:simul}

In this section, large scale simulations are performed to validate our algorithm for brain data.

To synthesize the data, the underlying true parameter values are fixed, 100 networks are randomly generated with the same $G = 5, R = 5, T = 4$ as the underlying true regulatory relationships. For each of the 100 parameter settings, one dataset with 20 observations at each stage is generated from the proposed model. And other experimental settings are given in Appendix~\ref{a1} in detail.

For comparison, other three simple methods, denoted as Pearson1, Pearson2 and Pearson3, respectively, through Pearson correlation coefficient are adopted and its basic steps are described in Algorithm~\ref{pearson123}. Pearson1 only use the observation dataset to estimate the networks, but Pearson2 and Pearson3 conduct missing data imputation firstly before estimating the networks, moreover, not only imputing $\times$, but $\bigcirc$ in Table\ref{dataexample}. For Pearson2, the missing data is estimated by the mean value of its corresponding observations, for instance, the estimators of missing $D_{g_i r_j,t'}^{e,t}$, $t' \ne t$, for all $e = 1, \cdots, n_t$ and $t = 1, \cdots, T$ are the same and equal to the mean value of the observations $\boldsymbol{D}_{g_i r_j,t'}^{t'}$. For Pearson3, we obtain the complete dataset by Random Forest Missing Data Algorithm \citep{Feitangmissforest}. Table~\ref{brainsimulation} shows the simulation results.

There are four evaluation indexes adopted to assess the performances. The `Detection' represents the ratio of number of regulatory relationships detected to number of regulatory relationships in true networks. The `Recall' represents the ratio of number of regulatory relationships correctly detected to number of regulatory relationships in true networks. The `Error' represents the ratio of total learning error (false positives plus false negatives) to number of regulatory relationships in true networks. The '$F_1$ score' represents the ratio $2 \cdot \frac{\text{Precision} \cdot \text{Recall}}{\text{Precision} + \text{Recall}}$.

As shown in Table~\ref{brainsimulation}, no matter what the index, no matter what the stage, the proposed approach is far better than other methods. To be more specific, our `Detection's are closer to 1 than the others and our `Recall's are superior to all others indicating that our method not only is very accurate, also have fewer false positive. Thus, our `Error's are much smaller and `$F_1$ score's are much larger than all others. However, relatively speaking, our method does not perform very well during the first stage transition. This phenomenon is understandable and explicable. During each stage transition, both its corresponding observations and missing data are used to learn its network. For the first stage transition, the missing data accounts for $(T-1) \times 100$\% and they are furthest from their corresponding observations, making them the most difficult to estimate, thus, the missing data affects the estimation accuracy to some extent.
\begin{algorithm}
	\caption{Algorithm for Pearson1, Pearson2 and Pearson3}
	\label{pearson123}
	\begin{algorithmic}[1]
		\REQUIRE ~~\\ 
		If Pearson1, input the observation $\boldsymbol{D}_{t}^{t}, t=1,\cdots,T$;\\
		If Pearson2 or Pearson 3, input the complete dataset $\boldsymbol{D}$ after imputation
		\ENSURE ~~\\ 
		The regulatory networks
		
		\FOR{$t=2;t\le T; t++$}
		\FOR{$g_i=1; g_i \le G; g_i++$}
		\FOR{$r_j=1; r_j \le R; r_j++$}
		\STATE If Pearson1, calculate Pearson correlation coefficients between $(\boldsymbol{D}_{g_i r_j,t-1}^{t-1}, \boldsymbol{D}_{g_i r_j,t}^t)$ and $(\boldsymbol{D}_{g_m r_n,t-1}^{t-1}, \boldsymbol{D}_{g_i r_j,t}^t)$ for all $(g_m, r_n) \ne (g_i, r_j)$;\\
		If Pearson2 or Pearson 3, calculate Pearson correlation coefficients between $\{ \boldsymbol{D}_{g_i r_j,t}^{t'}; t' = 1, \cdots, T \}$ and $\{ \boldsymbol{D}_{g_m r_n,t}^{t'}; t' = 1, \cdots, T \}$ for all $(g_m, r_n) \ne (g_i, r_j)$
		\IF{all correlation coefficients are smaller than 0.5}
		\STATE $(g_i, r_j)$ is not regulated during $(t-1, t)$
		\ELSE
		\STATE $(g_i, r_j)$ is regulated by $(g_m, r_n)$ with largest absolute value of correlation coefficient during $(t-1, t)$
		\ENDIF
		\ENDFOR
		\ENDFOR
		\ENDFOR
	\end{algorithmic}
\end{algorithm}

\begin{table}
	\centering
	\caption[Results of simulations for detecting regulatory relationships]{Results of simulations for detecting regulatory relationships. Each row represents one evaluation index during one stage transition or the whole transition and each column indicates one method. The values in brackets are the variations of the corresponding indexes.}
	\scalebox{0.9}{
		\begin{tabular}{c|c||c|c|c|c}
			\multicolumn{2}{c||}{ } & \multicolumn{1}{c|}{Pearson 1} & \multicolumn{1}{c|}{Pearson 2} & \multicolumn{1}{c|}{Pearson 3} & \multicolumn{1}{c}{Proposed} \\
			\hline
			\multirow{8}*{Stage 1 $\rightarrow$ 2}
			& \multirow{2}*{Detection}& \multirow{2}*{1.1857 (0.4228)}& \multirow{2}*{4.1251 (9.7894)}& \multirow{2}*{4.0321 (9.1675)}& \multirow{2}*{1 (0.023)} \\
			& & & & & \\
			\cline{2-6}
			& \multirow{2}*{Recall}& \multirow{2}*{0.0353 (0.012)}& \multirow{2}*{0.0058 (0.0019)}& \multirow{2}*{0.0058 (0.0019)}& \multirow{2}*{0.5239 (0.1237)} \\
			& & & & & \\
			\cline{2-6}
			& \multirow{2}*{Error}& \multirow{2}*{1.3004 (0.4182)}& \multirow{2}*{4.9458 (9.9604)}& \multirow{2}*{4.8646 (9.3752)}& \multirow{2}*{0.4875 (0.1471)} \\
			& & & & & \\
			\cline{2-6}
			& \multirow{2}*{$F_1$ score}& \multirow{2}*{0.0476 (0.0209)}& \multirow{2}*{0.0036 (7e-04)}& \multirow{2}*{0.004 (9e-04)}& \multirow{2}*{0.6083 (0.1264)} \\
			& & & & & \\
			\hline
			\hline
			\multirow{8}*{Stage 2 $\rightarrow$ 3}
			& \multirow{2}*{Detection}& \multirow{2}*{1.3238 (0.8217)}& \multirow{2}*{6.4895 (20.3016)}& \multirow{2}*{6.4597 (20.863)}& \multirow{2}*{0.9932 (0.0041)} \\
			& & & & & \\
			\cline{2-6}
			& \multirow{2}*{Recall}& \multirow{2}*{0.0258 (0.009)}& \multirow{2}*{0.0103 (0.0039)}& \multirow{2}*{0.0103 (0.0039)}& \multirow{2}*{0.7199 (0.1)} \\
			& & & & & \\
			\cline{2-6}
			& \multirow{2}*{Error}& \multirow{2}*{1.5181 (0.8867)}& \multirow{2}*{7.1702 (20.3729)}& \multirow{2}*{7.144 (20.9782)}& \multirow{2}*{0.2824 (0.1038)} \\
			& & & & & \\
			\cline{2-6}
			& \multirow{2}*{$F_1$ score}& \multirow{2}*{0.0308 (0.0118)}& \multirow{2}*{0.0037 (5e-04)}& \multirow{2}*{0.0038 (5e-04)}& \multirow{2}*{0.784 (0.0899)} \\
			& & & & & \\
			\hline
			\hline
			\multirow{8}*{Stage 3 $\rightarrow$ 4}
			& \multirow{2}*{Detection}& \multirow{2}*{1.9853 (1.5127)}& \multirow{2}*{6.6772 (23.1663)}& \multirow{2}*{6.5436 (22.1384)}& \multirow{2}*{1.1044 (0.1947)} \\
			& & & & & \\
			\cline{2-6}
			& \multirow{2}*{Recall}& \multirow{2}*{0.0537 (0.0158)}& \multirow{2}*{0.0088 (0.0023)}& \multirow{2}*{0.0098 (0.0025)}& \multirow{2}*{0.7326 (0.1107)} \\
			& & & & & \\
			\cline{2-6}
			& \multirow{2}*{Error}& \multirow{2}*{2.1129 (2.0116)}& \multirow{2}*{7.2945 (22.5546)}& \multirow{2}*{7.1683 (21.699)}& \multirow{2}*{0.3884 (0.3237)} \\
			& & & & & \\
			\cline{2-6}
			& \multirow{2}*{$F_1$ score}& \multirow{2}*{0.0578 (0.0189)}& \multirow{2}*{0.004 (5e-04)}& \multirow{2}*{0.0045 (5e-04)}& \multirow{2}*{0.762 (0.0942)} \\
			& & & & & \\
			\hline
			\hline
			\multirow{8}*{Total}
			& \multirow{2}*{Detection}& \multirow{2}*{1.4324 (0.1251)}& \multirow{2}*{4.4902 (2.0268)}& \multirow{2}*{4.4241 (2.0042)}& \multirow{2}*{1.008 (0.0097)} \\
			& & & & & \\
			\cline{2-6}
			& \multirow{2}*{Recall}& \multirow{2}*{0.0414 (0.0041)}& \multirow{2}*{0.0092 (0.0011)}& \multirow{2}*{0.0099 (0.0012)}& \multirow{2}*{0.6662 (0.0331)} \\
			& & & & & \\
			\cline{2-6}
			& \multirow{2}*{Error}& \multirow{2}*{1.4746 (0.128)}& \multirow{2}*{5.1945 (2.0181)}& \multirow{2}*{5.1311 (2.0082)}& \multirow{2}*{0.3551 (0.04)} \\
			& & & & & \\
			\cline{2-6}
			& \multirow{2}*{$F_1$ score}& \multirow{2}*{0.0533 (0.0071)}& \multirow{2}*{0.0038 (2e-04)}& \multirow{2}*{0.0041 (2e-04)}& \multirow{2}*{0.7759 (0.0221)} \\
			& & & & & \\
		\end{tabular}
	}
	\label{brainsimulation}
\end{table}

\section{Real Data Analyses}\label{sec:real}

We apply our method on a real dataset about genome-wide gene expression provided by \citep{haroutunian2009transcriptional}. The brain is divided into 19 regions and the progress of AD is stratified into 4 stages according to the density of cerebrocortical neuritic plaque. The higher the stage, the more severe the disease. A microarray analysis in human postmortem specimens was conducted by Gene Logic Inc. using Affymetrix. Data from 992 samples were normalized using MAS 5.0 algorithms and $\text{GX}^{\text{TM}}$ Explorer v.2.0 \citep{Katsel2005Variations}. The brain dataset handling procedures are described in detail by \citep{haroutunian2009transcriptional} and sample distributions of the data between different regions are shown in Appendix~\ref{a2}. These are the 19 regions: Anterior Cingulate (AC), Caudate Nucleus (CN), Dorsolateral Prefrontal Cortex (DPC), Frontal Pole (FP), Hippocampus (HIP), Inferior Frontal Gyrus (IFG), Inferior Temporal Gyrus (ITG), Middle Temporal Gyrus (MTG), Occipital Visual Cortex (OVC), Parahippocampal Gyrus (ParaG), Posterior Cingulate Cortex (PCC), Precentral Gyrus (PreG), Prefrontal Cortex (PC), Putamen (PUT), Superior Parietal Lobule (SPL), Superior Temporal Gyrus (STG), Temporal Pole (TP), Amygdala (Amyg), Nucleus Accumbens (NA).

In the previous simulations with 5 genes, 5 regions and 4 stages, it takes 6.182 hours for 10000 MCMC iterations. Because of the property of MCMC and the exponentially increasing of the number of possible regulatory relationships with the numbers of genes and regions, our algorithm may take a pretty long time when the number of genes or regions is very large. Thus, we suggest practical approaches to use our MCMC algorithm under are two situations to perform the real data analysis. 

The first situation is that the numbers of genes and regions that we're interested in are not very large. In this case, we can conduct one MCMC experiment to get the result. In this experiment, specific regions and genes are selected to demonstrate the rationality of our method. Five regions are selected: PC, HIP, MTG, STG, PG. PC is part of the frontal lobe, and it contributes to working memory \citep{boisgueheneuc2006functions}. HIP is associated with short-term, long-term memory and spatial memory \citep{suthana2012memory}. MTG, STG and PG are the three regions with the greatest changes in gene expression \citep{haroutunian2009transcriptional}. Fifteen genes are selected: $PHYHD1$, $MYO5C$, $ENPP2$, $MOG$, $GPR37$, $LIPA$, $MAG$, $PSEN1$, $TF$, $BIN1$, $CR1$, $SORL1$, $CASS4$, $CLU$, $PICALM$. These genes can be divided into three categories. It is likely that functional failure of $PHYHD1$ and $MYO5C$ could lead to AD development \citep{miyashita2014genes}. Seven genes, $ENPP2$, $MOG$, $GPR37$, $LIPA$, $MAG$, $PSEN1$, $TF$, are in the same module which is related to disease progression \citep{MillerA}. And the last six genes are associated with AD in genome-wide significance \citep{lambert2013meta}. In this real data analysis with 15 genes and 5 regions,  it takes 21.133 hours for 10000 MCMC iterations. The results are shown in Table~\ref{brainrealdata}.
\begin{table}
	\centering
	\caption[Results of real data for detecting regulatory relationships]{Results of real data for detecting regulatory relationships. Each $g_i, r_j - g_m, r_n$ represents one detected regulatory relationship and means that gene $g_i$ in region $r_j$ is regulated by gene $g_m$ in region $r_n$ during the corresponding stage transition, and the value in bracket is the proportion of this regulatory relationship in its corresponding MCMC samples. The indexes of regions, 1 to 5, correspond to PC, HIP, MTG, STG and PG in turn. The indexes of genes, 1 to 15, correspond to $PHYHD1$, $MYO5C$, $ENPP2$, $MOG$, $GPR37$, $LIPA$, $MAG$, $PSEN1$, $TF$, $BIN1$, $CR1$, $SORL1$, $CASS4$, $CLU$, $PICALM$ in turn.}
	\begin{tabular}{c||c}
		Stage transition & Detected regulatory relationships\\
		\hline
		Stage 1 $\rightarrow$ 2 & $\begin{matrix} \\ 3,2 - 5,2 (21.50\%) & 4,2 - 8,2 (21.60\%) & 4,5 - 7,5 (18.40\%)\\ 6,5 - 4,5 (38.64\%) \\ \\ \end{matrix}$\\
		\hline
		Stage 2 $\rightarrow$ 3 & $\begin{matrix} \\  \\ \end{matrix}$\\
		\hline
		Stage 3 $\rightarrow$ 4 & $\begin{matrix} \\ 2,5 - 13,4 (26.59\%) & 4,3 - 11,1 (21.89\%) & 4,5 - 11,1 (18.06\%)\\ 5,2 - 13,4 (15.43\%) & 9,3 - 12,1 (25.30\%) & 11,4 - 10,4 (21.12\%) \\ 12,1 - 11,5 (24.19\%) & 13,4 - 2,5 (20.58\%) & 15,1 - 13,5 (21.75\%) \\ \\ \end{matrix}$\\
	\end{tabular}
	\label{brainrealdata}
\end{table}

In Table~\ref{brainrealdata}, we can find that, during the first stage transition, all detected regulatory relationships exist within the same module mentioned above. During the last stage transition,  genes in the third category are involved most, which confirms their significance in relation to AD. Thus, this real data analysis shows that our method can reach conclusions that align well with existing biological knowledge.

The another situation is that the numbers of genes or regions are so large that one MCMC experiment may not handle them. The following procedure, Algorithm~\ref{procedure}, is proposed to detect the regulatory relationships with wide ranges of genes or regions. This algorithm is essentially a dimensionality reduction method, but genes and regions may overlap in different MCMC, thus, we can try to ensure that every possible situation maybe considered if the MCMC is used enough times. Based on the computational times in the simulation and first real data analysis, we can obviously find that this kind of procedure is pretty necessary for wide ranges of genes and regions, otherwise, one MCMC experiment might take too long to get the result. In this experiment, we adopt all regions and randomly select 100 genes to conduct the second real data analysis and we set $N_M = 10$, $G = 15$, $R = 5$. The full result is displayed in Appendix~\ref{a3} and its corresponding index tables of genes and regions are provided in Appendix~\ref{a4}. The total computational time is 219.345 hours.
\begin{algorithm}
	\caption{Algorithm for brain data with wide ranges of genes or regions}
	\label{procedure}
	\begin{algorithmic}[1]
		\REQUIRE ~~\\ 
		$N_G$ genes; $N_R$ regions; $T$ stages\\
		$G$: number of genes within one MCMC; $R$: number of regions within one MCMC; $N_M$: times of MCMC
		\ENSURE ~~\\ 
		The regulatory networks
		
		\FOR{$g_i=1; g_i \le N_G; g_i++$}
		\STATE Calculate the Pearson correlation coefficients between $\{ \boldsymbol{D}_{g_i r_j,t}^{t}; t = 1, \cdots, T, r_j = 1, \cdots, N_R \}$ and $\{ \boldsymbol{D}_{g_m r_n,t}^{t}; t = 1, \cdots, T, r_n = 1, \cdots, N_R \}$ for all $g_m \ne g_i$, denoted as $Cor_{g_i}^{g_m}$;
		\ENDFOR
		\FOR{$r_j=1; r_j \le N_R; r_j++$}
		\STATE Conduct the hypothesis test: $H_0: D_{g_i r_j,t}^t = D_{g_m r_n,t}^t$ $\textbf{VS}$ $H_1: D_{g_i r_j,t}^t \ne D_{g_m r_n,t}^t$ for all $(g_i, r_j)$, $t \ge 2$ and denote its p-value as $P_{g_i r_j}^t$;
		\ENDFOR
		\FOR{$m_t = 1; m_t \le N_M; m_t++$}
		\STATE Sample $G$ genes from the whole $N_G$ genes without replacement according to the probability of $g_i$ satisfying: $P(g_i) \propto \sum_{g_m \ne g_i}Cor_{g_i}^{g_m}$;
		\STATE Sample $R$ regions from the whole $N_R$ regions without replacement according to the probability of $r_j$ satisfying: $P(r_j) \propto \#\{P_{g_i r_j}^t <= 0.05; t = 2, \cdots, T, g_i = 1, \cdots, N_G\}$;
		\STATE Input the $G$ genes and $R$ regions into Algorithm~\ref{algorithm};
		\ENDFOR
	\end{algorithmic}
\end{algorithm}

\section{Conclusion}\label{sec:con}

In this study, a full Bayesian framework is proposed to detect gene regulatory relationships of AD brain during each stage transition. Simulations are conducted to validate the statistical power of our algorithm. Moreover, a real data analysis shows that our method can capture the gene regulatory relationships among this complex brain data.

The majority of other methods to identify gene regulatory relationships or analysis transcriptional changes are processing region by region of brain, stage by stage of AD, and then compare these sub-results to detect relationships or changes. However, such approach would lack the integrity of regions and the coherence of stages. We present an overall statistical model of regulatory relationships in AD, thus, we can deal with gene, region and stage, these three dimensions, simultaneously to study genome-wide gene expression dynamics.

Our model and algorithm can also be applied to other areas. For example, dynamic network learning with time dimension, it can also degenerate into static network learning.

Even so, there are still some problems to be solved. For instance, the regulatory mechanism in the model is relatively simple, and more complex cases need further research. The relationship between regions is not reflected in the model. And the computational power of algorithms also needs to be improved. These will be the focus of our future work.

\bibliographystyle{agsm}
\bibliography{bib}

\section*{Appendix}\label{app}
\appendix
\setcounter{table}{0}
\setcounter{figure}{0}
\setcounter{equation}{0}
\renewcommand{\thesubsection}{A.\arabic{subsection}}
\renewcommand{\thetable}{A.\arabic{table}}
\renewcommand{\thefigure}{A.\arabic{figure}}
\renewcommand{\theequation}{A.\arabic{equation}}

\subsection{Experimental Settings for Brain Data}\label{a1}

We used the following hyper-parameters for the prior distributions:
\begin{gather*}
	c_{g_i r_j} = 5, d_{g_i r_j} = 0.5 \\
	c_2 = 0, d_2 = 0.5 \\
	p_i = 3, q_i = 2, \text{   for } i = 1,2\\
	\boldsymbol{\alpha}_{g_i r_j,t-1,t}^{g_m r_n} = \begin{pmatrix} 1 \\ 1 \end{pmatrix}, \boldsymbol{V}_{g_i r_j,t-1,t}^{g_m r_n} = \begin{pmatrix} 1 & 0 \\ 0 & 1 \end{pmatrix}\\
	v = 2, \lambda = 0.05
\end{gather*}
for all possible $(g_i, r_j)$, $(g_m, r_n)$ and $t$.

The matrix of probability function of alternative operations to the modes is given as follows:
\[ \left(\left. \begin{tabular}{c|c|c|c}
	Model Type & Add & Delete & Swap \\
	\hline
	No Relationship & 1 & 0 & 0 \\
	\hline
	All Regulated & 0 & 0.8 & 0.2 \\
	\hline
	Other Case & 0.3 & 0.4 & 0.3
\end{tabular}\right.\right) \]

\subsection{Sample Distributions for Brain Data}\label{a2}
\begin{figure}
	\centering
	\includegraphics[scale=0.25]{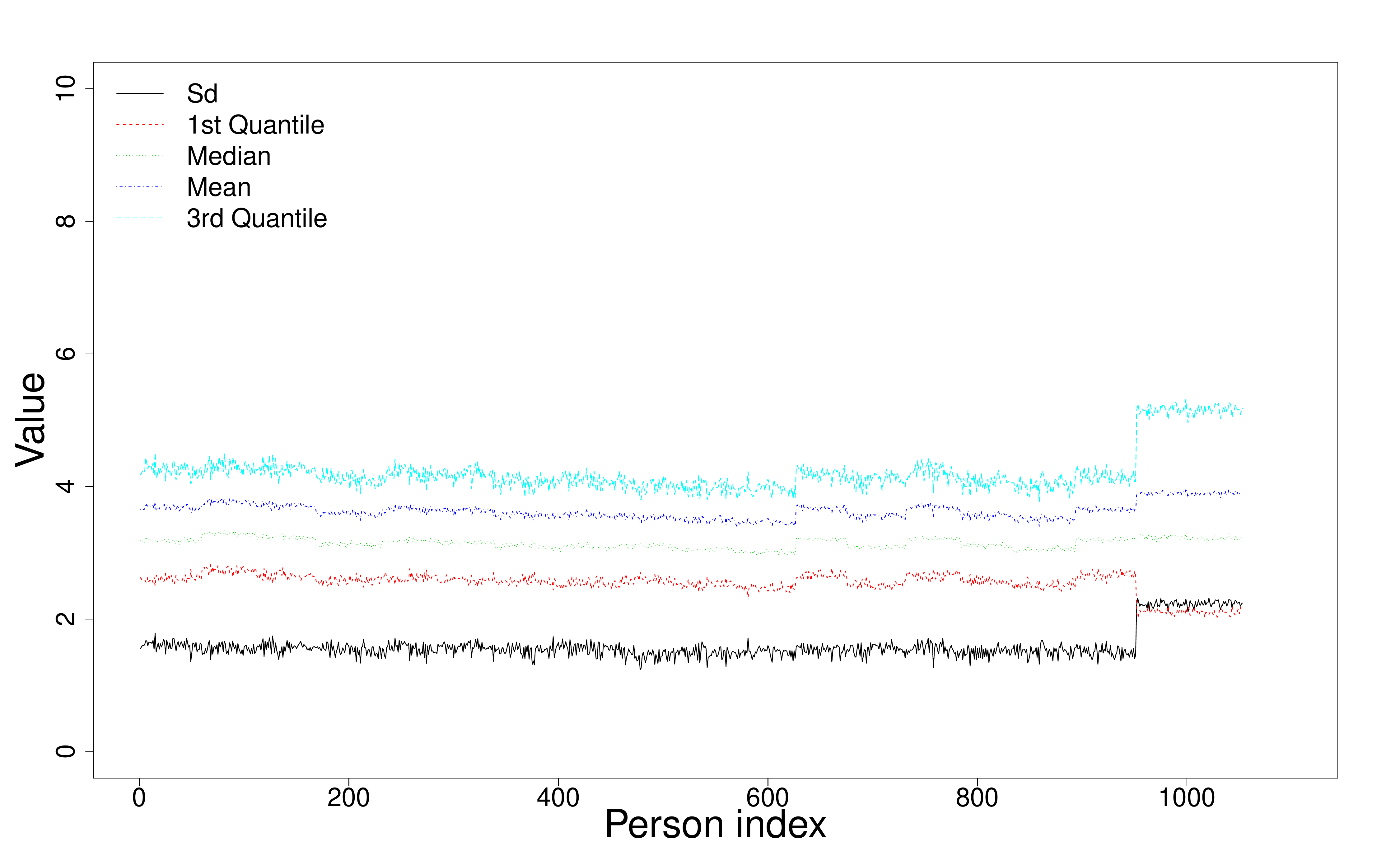}
	\caption[Sample distribution characteristics of All Persons for brain data]{Sample distribution characteristics of All Persons for brain data.}
	\label{brainrealdata1}
\end{figure}

\begin{figure}
	\centering
	\includegraphics[scale=0.25]{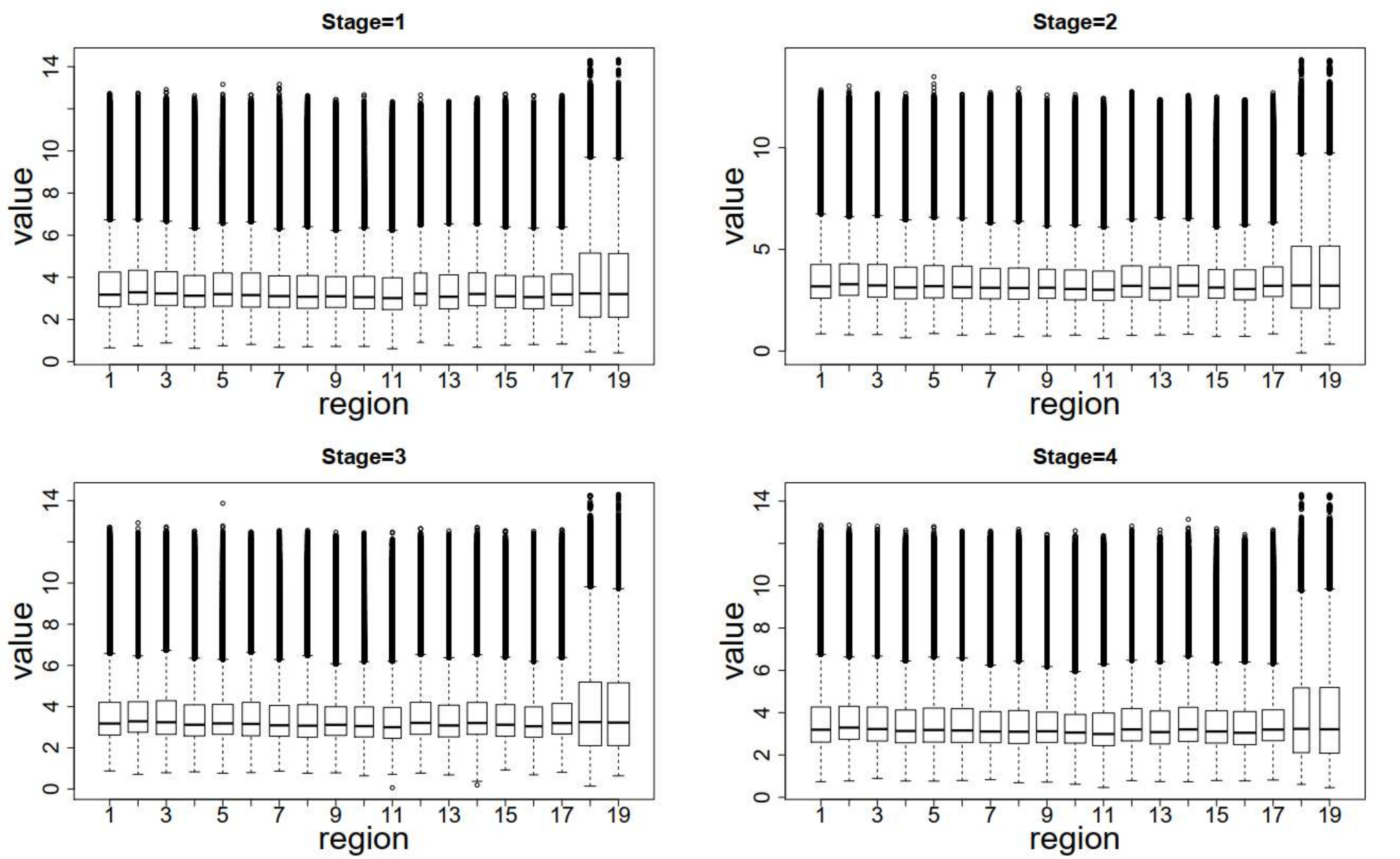}
	\caption[Box plots of the sample distributions in different regions for brain data]{Box plots of the sample distributions in different regions for brain data. Each box plot shows the sample distribution corresponding to a same region.}
	\label{brainrealdata2}
\end{figure}

For the gene expression of each person, we calculate its sample distribution characteristics, i.e., standard deviation, median, mean, 1st and 3rd quantiles, based on the observations of all genes. For each region and each stage, we put the observations of all genes and its all corresponding persons together to plot its box plot. The results are shown in Figure~\ref{brainrealdata1} and Figure~\ref{brainrealdata2}, respectively. We can find that the distributions of gene expression in different gene chips are similar after normalization.


\subsection{Results of Brains data with Wide Range of Genes and Regions}\label{a3}

\begin{table}
	\centering
	\caption{Results of real data for detecting regulatory relationships with wide range of genes and regions. Each $g_i, r_j - g_m, r_n$ represents one detected regulatory relationship and means that gene $g_i$ in region $r_j$ is regulated by gene $g_m$ in region $r_n$ during the corresponding stage transition, and the value in bracket is the proportion of this regulatory relationship in its corresponding MCMC samples.}
	\begin{tabular}{c||c}
		Stage transition & Detected regulatory relationships\\
		\hline
		Stage 1 $\rightarrow$ 2 & $\begin{matrix} \\ 1,8 - 92,3 (50.93\%) &  1,18 - 26,10 (48.21\%) &  1,19 - 92,8 (50.00\%) \\  5,6 - 16,16 (23.33\%) &  9,14 - 11,11 (31.18\%) &  11,1 - 37,5 (39.17\%) \\  12,16 - 24,16 (46.41\%) &  15,12 - 98,1 (48.35\%) &  15,19 - 36,12 (37.52\%) \\  18,16 - 18,12 (47.72\%) &  20,8 - 92,3 (50.00\%) &  20,18 - 76,16 (48.24\%) \\  27,4 - 68,3 (50.86\%) &  27,7 - 94,18 (40.46\%) &  29,1 - 53,1 (42.16\%) \\  29,12 - 53,17 (39.37\%) &  33,3 - 23,19 (27.00\%) &  33,12 - 92,8 (50.15\%) \\  36,12 - 92,1 (37.87\%) &  36,19 - 62,8 (45.55\%) &  39,19 - 71,19 (18.54\%) \\  40,1 - 40,19 (50.00\%) &  43,16 - 50,18 (39.62\%) &  43,18 - 94,7 (27.58\%) \\  43,18 - 66,10 (50.21\%) &  43,19 - 96,8 (28.52\%) &  48,19 - 92,8 (50.00\%) \\  50,7 - 87,18 (49.27\%) &  51,18 - 49,8 (50.77\%) &  53,17 - 64,15 (25.04\%) \\  57,18 - 66,10 (18.46\%) &  61,18 - 74,7 (23.89\%) &  62,1 - 98,1 (50.00\%) \\  62,12 - 40,3 (39.49\%) &  62,17 - 76,17 (34.35\%) &  63,7 - 41,3 (49.30\%) \\  63,17 - 63,12 (43.52\%) &  63,18 - 94,7 (50.17\%) &  64,16 - 58,16 (21.50\%) \\  65,18 - 65,2 (19.05\%) &  68,16 - 33,4 (44.89\%) &  70,19 - 71,19 (31.58\%) \\  74,18 - 31,16 (29.88\%) &  81,18 - 10,16 (31.82\%) &  88,2 - 61,18 (21.99\%) \\  91,11 - 73,11 (50.00\%) &  92,8 - 98,1 (50.76\%) &  92,12 - 33,3 (50.00\%) \\  96,3 - 33,12 (28.08\%) &  96,18 - 63,16 (41.63\%) &  98,19 - 20,1 (56.54\%) \\  99,18 - 66,11 (22.23\%) \\ \\ \end{matrix}$\\
		\hline
		Stage 2 $\rightarrow$ 3 & $\begin{matrix} \\ 3,16 - 88,16 (50.06\%) &  9,6 - 11,1 (49.82\%) &  11,14 - 56,19 (44.36\%) \\  16,6 - 11,2 (38.39\%) &  16,14 - 11,14 (39.03\%) &  16,15 - 53,12 (40.25\%) \\  26,11 - 61,14 (36.41\%) &  31,1 - 18,16 (49.66\%) &  32,2 - 91,11 (33.95\%) \\  43,10 - 49,1 (37.88\%) &  43,18 - 68,16 (38.79\%) &  43,18 - 97,1 (36.58\%) \\  44,6 - 69,16 (48.20\%) &  44,9 - 65,13 (50.03\%) &  45,11 - 20,14 (39.04\%) \\  50,10 - 66,18 (34.20\%) &  51,4 - 13,1 (31.22\%) &  51,9 - 55,9 (30.28\%) \\  51,19 - 41,9 (50.00\%) &  53,1 - 53,16 (38.19\%) &  56,1 - 65,10 (50.00\%) \\  56,19 - 11,14 (39.12\%) &  58,16 - 73,15 (50.00\%) &  61,6 - 11,16 (48.47\%) \\  61,16 - 88,16 (50.00\%) &  63,1 - 18,16 (48.08\%) &  63,10 - 65,13 (50.00\%) \\  67,16 - 63,1 (50.00\%) &  70,6 - 13,19 (34.14\%) &  70,14 - 11,19 (35.11\%) \\  72,18 - 49,1 (38.42\%) &  76,17 - 53,16 (46.30\%) &  78,19 - 71,16 (42.04\%) \\  80,6 - 56,6 (27.01\%) &  81,6 - 60,16 (33.44\%) &  81,14 - 70,16 (40.66\%) \\  82,10 - 88,16 (50.07\%) &  82,16 - 22,14 (29.40\%) &  87,1 - 3,16 (50.00\%) \\  89,13 - 73,2 (35.29\%) &  90,6 - 11,1 (50.41\%) &  96,1 - 36,3 (43.64\%) \\  97,11 - 57,1 (25.77\%) &  99,10 - 49,1 (50.02\%) &  99,11 - 97,8 (20.95\%) \\ \\ \end{matrix}$ \\
	\end{tabular}
	\label{brainrealdata21}
\end{table}
\begin{table}
	\centering
	\caption{Results of real data for detecting regulatory relationships with wide range of genes and regions (continued Figure~\ref{brainrealdata21}).}
	\scalebox{0.9}{
		\begin{tabular}{c||c}
			Stage transition & Detected regulatory relationships\\
			\hline
			Stage 3 $\rightarrow$ 4 & $\begin{matrix} \\ 1,12 - 44,3 (16.00\%) &  3,16 - 88,16 (26.04\%) &  5,6 - 22,6 (11.19\%) \\  9,2 - 56,2 (13.76\%) &  9,11 - 11,11 (22.29\%) &  11,2 - 20,2 (26.35\%) \\  11,10 - 88,2 (14.86\%) &  13,18 - 81,18 (35.70\%) &  16,17 - 54,12 (18.45\%) \\  18,5 - 20,2 (28.85\%) &  18,15 - 69,17 (15.00\%) &  18,16 - 63,17 (29.49\%) \\  20,1 - 36,1 (16.40\%) &  20,2 - 18,5 (26.70\%) &  20,2 - 88,13 (09.92\%) \\  20,12 - 36,3 (31.99\%) &  20,19 - 36,12 (51.27\%) &  21,1 - 62,17 (11.26\%) \\  21,15 - 54,15 (20.06\%) &  21,17 - 58,12 (22.19\%) &  21,19 - 26,4 (09.48\%) \\  23,1 - 48,12 (31.45\%) &  24,5 - 31,7 (18.42\%) &  24,13 - 88,13 (16.91\%) \\  24,16 - 65,16 (17.02\%) &  24,16 - 61,16 (11.32\%) &  25,2 - 20,10 (12.57\%) \\  25,11 - 20,14 (14.65\%) &  27,3 - 41,4 (10.17\%) &  27,7 - 91,3 (11.88\%) \\  29,15 - 16,15 (18.37\%) &  29,16 - 53,1 (28.94\%) &  31,7 - 24,5 (21.97\%) \\  31,18 - 69,16 (14.99\%) &  32,2 - 88,2 (16.74\%) &  32,13 - 88,2 (19.72\%) \\  36,1 - 20,1 (18.31\%) &  36,3 - 20,12 (29.96\%) &  36,12 - 20,19 (52.79\%) \\  36,19 - 48,1 (11.61\%) &  37,1 - 69,1 (27.17\%) &  37,6 - 69,16 (28.83\%) \\  39,10 - 89,11 (13.92\%) &  39,11 - 11,14 (15.03\%) &  41,16 - 13,19 (06.42\%) \\  43,10 - 49,1 (16.14\%) &  43,18 - 89,18 (19.55\%) &  44,1 - 96,3 (31.38\%) \\  44,9 - 64,10 (08.35\%) &  45,11 - 53,14 (13.40\%) &  45,15 - 62,12 (40.92\%) \\  48,12 - 23,1 (28.48\%) &  49,1 - 49,8 (18.15\%) &  50,10 - 89,11 (24.62\%) \\  51,10 - 89,1 (16.20\%) &  53,1 - 29,16 (27.21\%) &  53,2 - 56,14 (14.06\%) \\  53,12 - 18,15 (27.94\%) &  56,1 - 71,16 (09.11\%) &  56,9 - 88,1 (18.58\%) \\  56,13 - 45,11 (11.97\%) &  58,16 - 63,3 (10.01\%) &  58,16 - 63,12 (50.00\%) \\  59,9 - 24,1 (29.02\%) &  59,10 - 63,13 (13.92\%) &  59,16 - 63,10 (49.07\%) \\  60,10 - 59,10 (16.00\%) &  60,16 - 24,16 (14.86\%) &  61,14 - 9,13 (20.25\%) \\  61,16 - 64,9 (20.86\%) &  62,8 - 20,3 (16.84\%) &  62,12 - 45,15 (44.75\%) \\  62,17 - 21,12 (22.81\%) &  63,1 - 67,16 (43.92\%) &  63,10 - 59,16 (49.44\%) \\  63,12 - 58,16 (57.33\%) &  64,1 - 69,12 (13.40\%) &  64,15 - 63,16 (09.59\%) \\  65,9 - 87,10 (19.58\%) &  65,10 - 71,16 (09.40\%) &  66,11 - 89,8 (18.14\%) \\  67,1 - 87,10 (22.92\%) &  67,9 - 56,13 (11.65\%) &  67,16 - 63,1 (47.88\%) \\  68,16 - 76,7 (19.25\%) &  69,1 - 37,1 (26.98\%) &  69,1 - 66,10 (25.89\%) \\  71,4 - 13,4 (08.44\%) &  72,1 - 49,11 (15.47\%) &  72,8 - 49,8 (13.95\%) \\  72,10 - 49,10 (15.05\%) &  73,2 - 20,10 (19.41\%) &  73,10 - 56,11 (18.61\%) \\  73,17 - 69,17 (31.77\%) &  76,4 - 58,7 (13.88\%) &  76,7 - 68,16 (19.58\%) \\  76,7 - 13,2 (19.82\%) &  78,19 - 13,4 (10.11\%) &  80,8 - 57,10 (20.02\%) \\  80,10 - 97,11 (45.97\%) &  81,18 - 13,18 (35.14\%) &  82,6 - 22,2 (07.04\%) \\  82,10 - 87,9 (05.76\%) &  82,19 - 13,1 (06.66\%) &  87,1 - 65,10 (10.02\%) \\  87,10 - 67,1 (25.84\%) &  88,1 - 64,9 (17.45\%) &  88,2 - 32,2 (18.04\%) \\  88,10 - 91,2 (23.54\%) &  88,13 - 24,13 (34.94\%) &  88,14 - 61,14 (10.14\%) \\  88,16 - 56,9 (15.43\%) &  89,10 - 90,10 (28.59\%) &  89,18 - 50,8 (34.24\%) \\  90,1 - 50,10 (11.29\%) &  91,10 - 39,10 (18.97\%) &  96,3 - 44,1 (27.81\%) \\  97,11 - 80,10 (46.07\%) &  98,8 - 62,3 (16.72\%) &  99,8 - 80,10 (16.76\%) \\  99,10 - 49,8 (31.32\%) &  99,11 - 1,10 (32.62\%) \\ \end{matrix}$\\
	\end{tabular}}
	\label{brainrealdata22}
\end{table}

\subsection{The indexes of Genes and Regions in Brain Data}\label{a4}

\begin{table}
	\centering
	\caption{Index table of regions in Brain data.}
	\scalebox{0.9}{
		\begin{tabular}{c|c|c|c}
			\multirow{2}*{Index} &  \multirow{2}*{1} &  \multirow{2}*{2} &  \multirow{2}*{3}  \\ & & & \\ \hline \multirow{2}*{Region} &  \multirow{2}*{Anterior Cingulate} &  \multirow{2}*{Caudate Nucleus} &  \multirow{2}*{Dorsolateral Prefrontal Cortex}  \\ & & & \\ \hline \multirow{2}*{Index} &  \multirow{2}*{4} &  \multirow{2}*{5} &  \multirow{2}*{6}  \\ & & & \\ \hline \multirow{2}*{Region} &  \multirow{2}*{Frontal Pole} &  \multirow{2}*{Hippocampus} &  \multirow{2}*{Inferior Frontal Gyrus}  \\ & & & \\ \hline \multirow{2}*{Index} &  \multirow{2}*{7} &  \multirow{2}*{8} &  \multirow{2}*{9}  \\ & & & \\ \hline \multirow{2}*{Region} &  \multirow{2}*{Inferior Temporal Gyrus} &  \multirow{2}*{Middle Temporal Gyrus} &  \multirow{2}*{Occipital Visual Cortex}  \\ & & & \\ \hline \multirow{2}*{Index} &  \multirow{2}*{10} &  \multirow{2}*{11} &  \multirow{2}*{12}  \\ & & & \\ \hline \multirow{2}*{Region} &  \multirow{2}*{Parahippocampal Gyrus} &  \multirow{2}*{Posterior Cingulate Cortex} &  \multirow{2}*{Precentral Gyrus}  \\ & & & \\ \hline \multirow{2}*{Index} &  \multirow{2}*{13} &  \multirow{2}*{14} &  \multirow{2}*{15}  \\ & & & \\ \hline \multirow{2}*{Region} &  \multirow{2}*{Prefrontal Cortex} &  \multirow{2}*{Putamen} &  \multirow{2}*{Superior Parietal Lobule}  \\ & & & \\ \hline \multirow{2}*{Index} &  \multirow{2}*{16} &  \multirow{2}*{17} &  \multirow{2}*{18}  \\ & & & \\ \hline \multirow{2}*{Region} &  \multirow{2}*{Superior Temporal Gyrus} &  \multirow{2}*{Temporal Pole} &  \multirow{2}*{Amygdala}  \\ & & & \\ \hline \multirow{2}*{Index} &  \multirow{2}*{19}  & & \\ & & & \\ \hline \multirow{2}*{Region} &  \multirow{2}*{Nucleus Accumbens}  & & \\ & & & \\
	\end{tabular}}
\end{table}

\begin{table}
	\centering
	\caption{Index table of genes in Brain data.}
	\scalebox{0.8}{
		\begin{tabular}{c|c|c|c|c|c|c|c|c}
			\multirow{2}*{Index} &  \multirow{2}*{1} &  \multirow{2}*{2} &  \multirow{2}*{3} &  \multirow{2}*{4} &  \multirow{2}*{5} &  \multirow{2}*{6} &  \multirow{2}*{7} &  \multirow{2}*{8}  \\ & & & & & & & & \\ \hline \multirow{2}*{Region} &  \multirow{2}*{phyhd1} &  \multirow{2}*{myo5c} &  \multirow{2}*{enpp2} &  \multirow{2}*{mog} &  \multirow{2}*{gpr37} &  \multirow{2}*{lipa} &  \multirow{2}*{mag} &  \multirow{2}*{psen1}  \\ & & & & & & & & \\ \hline \multirow{2}*{Index} &  \multirow{2}*{9} &  \multirow{2}*{10} &  \multirow{2}*{11} &  \multirow{2}*{12} &  \multirow{2}*{13} &  \multirow{2}*{14} &  \multirow{2}*{15} &  \multirow{2}*{16}  \\ & & & & & & & & \\ \hline \multirow{2}*{Region} &  \multirow{2}*{tf} &  \multirow{2}*{bin1} &  \multirow{2}*{cr1} &  \multirow{2}*{sorl1} &  \multirow{2}*{cass4} &  \multirow{2}*{clu} &  \multirow{2}*{picalm} &  \multirow{2}*{ash2l}  \\ & & & & & & & & \\ \hline \multirow{2}*{Index} &  \multirow{2}*{17} &  \multirow{2}*{18} &  \multirow{2}*{19} &  \multirow{2}*{20} &  \multirow{2}*{21} &  \multirow{2}*{22} &  \multirow{2}*{23} &  \multirow{2}*{24}  \\ & & & & & & & & \\ \hline \multirow{2}*{Region} &  \multirow{2}*{lyn} &  \multirow{2}*{tchp} &  \multirow{2}*{aqp4} &  \multirow{2}*{prickle3} &  \multirow{2}*{kiaa0494} &  \multirow{2}*{ecscr} &  \multirow{2}*{znf282} &  \multirow{2}*{fxyd3}  \\ & & & & & & & & \\ \hline \multirow{2}*{Index} &  \multirow{2}*{25} &  \multirow{2}*{26} &  \multirow{2}*{27} &  \multirow{2}*{28} &  \multirow{2}*{29} &  \multirow{2}*{30} &  \multirow{2}*{31} &  \multirow{2}*{32}  \\ & & & & & & & & \\ \hline \multirow{2}*{Region} &  \multirow{2}*{ccl11} &  \multirow{2}*{gpr172a} &  \multirow{2}*{scg5} &  \multirow{2}*{c1orf107} &  \multirow{2}*{cdkn2a} &  \multirow{2}*{itgam} &  \multirow{2}*{map3k9} &  \multirow{2}*{pde9a}  \\ & & & & & & & & \\ \hline \multirow{2}*{Index} &  \multirow{2}*{33} &  \multirow{2}*{34} &  \multirow{2}*{35} &  \multirow{2}*{36} &  \multirow{2}*{37} &  \multirow{2}*{38} &  \multirow{2}*{39} &  \multirow{2}*{40}  \\ & & & & & & & & \\ \hline \multirow{2}*{Region} &  \multirow{2}*{ip6k2} &  \multirow{2}*{upp1} &  \multirow{2}*{sh3gl3} &  \multirow{2}*{sprr1a} &  \multirow{2}*{trappc9} &  \multirow{2}*{diras2} &  \multirow{2}*{hace1} &  \multirow{2}*{ddx3x}  \\ & & & & & & & & \\ \hline \multirow{2}*{Index} &  \multirow{2}*{41} &  \multirow{2}*{42} &  \multirow{2}*{43} &  \multirow{2}*{44} &  \multirow{2}*{45} &  \multirow{2}*{46} &  \multirow{2}*{47} &  \multirow{2}*{48}  \\ & & & & & & & & \\ \hline \multirow{2}*{Region} &  \multirow{2}*{upk1a} &  \multirow{2}*{capn1} &  \multirow{2}*{hs6st2} &  \multirow{2}*{cyb5d1} &  \multirow{2}*{phf6} &  \multirow{2}*{cwc27} &  \multirow{2}*{loc100507448} &  \multirow{2}*{nmi}  \\ & & & & & & & & \\ \hline \multirow{2}*{Index} &  \multirow{2}*{49} &  \multirow{2}*{50} &  \multirow{2}*{51} &  \multirow{2}*{52} &  \multirow{2}*{53} &  \multirow{2}*{54} &  \multirow{2}*{55} &  \multirow{2}*{56}  \\ & & & & & & & & \\ \hline \multirow{2}*{Region} &  \multirow{2}*{fgf8} &  \multirow{2}*{slco3a1} &  \multirow{2}*{rab14} &  \multirow{2}*{u2af1} &  \multirow{2}*{loc100506318} &  \multirow{2}*{ncrna00158} &  \multirow{2}*{c16orf46} &  \multirow{2}*{loc100506305}  \\ & & & & & & & & \\ \hline \multirow{2}*{Index} &  \multirow{2}*{57} &  \multirow{2}*{58} &  \multirow{2}*{59} &  \multirow{2}*{60} &  \multirow{2}*{61} &  \multirow{2}*{62} &  \multirow{2}*{63} &  \multirow{2}*{64}  \\ & & & & & & & & \\ \hline \multirow{2}*{Region} &  \multirow{2}*{chchd6} &  \multirow{2}*{tlr6} &  \multirow{2}*{nrg4} &  \multirow{2}*{c10orf95} &  \multirow{2}*{slc35e1} &  \multirow{2}*{loc100509703} &  \multirow{2}*{slc35a3} &  \multirow{2}*{hdac2}  \\ & & & & & & & & \\ \hline \multirow{2}*{Index} &  \multirow{2}*{65} &  \multirow{2}*{66} &  \multirow{2}*{67} &  \multirow{2}*{68} &  \multirow{2}*{69} &  \multirow{2}*{70} &  \multirow{2}*{71} &  \multirow{2}*{72}  \\ & & & & & & & & \\ \hline \multirow{2}*{Region} &  \multirow{2}*{nucb1} &  \multirow{2}*{dpf3} &  \multirow{2}*{ltn1} &  \multirow{2}*{nox3} &  \multirow{2}*{adam1} &  \multirow{2}*{ttll7} &  \multirow{2}*{shc4} &  \multirow{2}*{fam114a2}  \\ & & & & & & & & \\ \hline \multirow{2}*{Index} &  \multirow{2}*{73} &  \multirow{2}*{74} &  \multirow{2}*{75} &  \multirow{2}*{76} &  \multirow{2}*{77} &  \multirow{2}*{78} &  \multirow{2}*{79} &  \multirow{2}*{80}  \\ & & & & & & & & \\ \hline \multirow{2}*{Region} &  \multirow{2}*{cd160} &  \multirow{2}*{enpp1} &  \multirow{2}*{las1l} &  \multirow{2}*{nudt4} &  \multirow{2}*{cldn14} &  \multirow{2}*{cgrrf1} &  \multirow{2}*{kbtbd8} &  \multirow{2}*{aoc2}  \\ & & & & & & & & \\ \hline \multirow{2}*{Index} &  \multirow{2}*{81} &  \multirow{2}*{82} &  \multirow{2}*{83} &  \multirow{2}*{84} &  \multirow{2}*{85} &  \multirow{2}*{86} &  \multirow{2}*{87} &  \multirow{2}*{88}  \\ & & & & & & & & \\ \hline \multirow{2}*{Region} &  \multirow{2}*{loc401052} &  \multirow{2}*{phf1} &  \multirow{2}*{sh2b1} &  \multirow{2}*{sry} &  \multirow{2}*{loc100507619} &  \multirow{2}*{nedd9} &  \multirow{2}*{iqsec2} &  \multirow{2}*{rfwd2}  \\ & & & & & & & & \\ \hline \multirow{2}*{Index} &  \multirow{2}*{89} &  \multirow{2}*{90} &  \multirow{2}*{91} &  \multirow{2}*{92} &  \multirow{2}*{93} &  \multirow{2}*{94} &  \multirow{2}*{95} &  \multirow{2}*{96}  \\ & & & & & & & & \\ \hline \multirow{2}*{Region} &  \multirow{2}*{wdr72} &  \multirow{2}*{fezf2} &  \multirow{2}*{ebf4} &  \multirow{2}*{ncoa3} &  \multirow{2}*{egr3} &  \multirow{2}*{loc100128496} &  \multirow{2}*{pvt1} &  \multirow{2}*{fign}  \\ & & & & & & & & \\ \hline \multirow{2}*{Index} &  \multirow{2}*{97} &  \multirow{2}*{98} &  \multirow{2}*{99} &  \multirow{2}*{100}  & & & & \\ & & & & & & & & \\ \hline \multirow{2}*{Region} &  \multirow{2}*{tubb1} &  \multirow{2}*{loc283624} &  \multirow{2}*{bloc1s2} &  \multirow{2}*{plekhh2}  & & & & \\ & & & & & & & & \\
		\end{tabular}
	}
\end{table}

\end{document}